\input harvmac
\newcount\yearltd\yearltd=\year\advance\yearltd by 0

\noblackbox

\input epsf

\def\tilde{\widetilde}
\def\hat{\widehat}

\newcount\figno
\figno=0
\def\fig#1#2#3{
\par\begingroup\parindent=0pt\leftskip=1cm\rightskip=1cm\parindent=0pt
\baselineskip=11pt
\global\advance\figno by 1
\midinsert
\epsfxsize=#3
\centerline{\epsfbox{#2}}
\vskip 12pt
{\bf Fig.\ \the\figno: } #1\par
\endinsert\endgroup\par
}
\def\figlabel#1{\xdef#1{\the\figno}}
\def\encadremath#1{\vbox{\hrule\hbox{\vrule\kern8pt\vbox{\kern8pt
\hbox{$\displaystyle #1$}\kern8pt}
\kern8pt\vrule}\hrule}}

\def\half{{\textstyle{1\over2}}}

\def\inbar{\vrule height1.5ex width.4pt depth0pt}
\def\half{{1\over 2}}

 \def\ep{{\epsilon}}

 \def\frac#1#2{{#1\over #2}}

 \def\D{{\Delta}}
 \def\g{{\gamma}}

 \def\Ph{{\Phi }}
 
 \def\CN{{\cal N}}
 
 \def\p{\partial}

\def\IR{\relax{\rm I\kern-.18em R}}
\def\IC{\relax\hbox{$\inbar\kern-.3em{\rm C}$}}
\def\IZ{\relax\ifmmode\hbox{Z\kern-.4em Z}\else{Z\kern-.4em Z}\fi}

\lref\wittenhol{
  E.~Witten,
  ``Anti-de Sitter space and holography,''
  Adv.\ Theor.\ Math.\ Phys.\  {\bf 2}, 253 (1998)
  [arXiv:hep-th/9802150].
}

\lref\sundborg{
B.~Sundborg,
``The Hagedorn transition, deconfinement and $N = 4$ SYM theory,''
Nucl.\ Phys.\ B {\bf 573}, 349 (2000)
[arXiv:hep-th/9908001].
}
\lref\apostol{
Tom Apostol
}

\lref\myconf{
S.~Minwalla,
``Restrictions imposed by superconformal invariance on quantum field
theories,''
Adv.\ Theor.\ Math.\ Phys.\  {\bf 2}, 781 (1998)
[arXiv:hep-th/9712074].
}

\lref\AharonyTI{
O.~Aharony, S.~S.~Gubser, J.~M.~Maldacena, H.~Ooguri and Y.~Oz,
``Large N field theories, string theory and gravity,''
Phys.\ Rept.\  {\bf 323}, 183 (2000)
[arXiv:hep-th/9905111].
}

\lref\suvrat{
  A.~Barabanschikov, L.~Grant, L.~L.~Huang and S.~Raju,
  ``The spectrum of Yang Mills on a sphere,''
  arXiv:hep-th/0501063.}

\lref\MaldacenaRE{
J.~M.~Maldacena,
``The large N limit of superconformal field theories and supergravity,''
Adv.\ Theor.\ Math.\ Phys.\  {\bf 2}, 231 (1998)
[Int.\ J.\ Theor.\ Phys.\  {\bf 38}, 1113 (1999)]
[arXiv:hep-th/9711200].
}
\lref\bianchione{
  M.~Bianchi, J.~F.~Morales and H.~Samtleben,
  ``On stringy AdS(5) x S**5 and higher spin holography,''
  JHEP {\bf 0307}, 062 (2003)
  [arXiv:hep-th/0305052].
}
\lref\bianchithree{
  N.~Beisert, M.~Bianchi, J.~F.~Morales and H.~Samtleben,
  ``Higher spin symmetry and N = 4 SYM,''
  JHEP {\bf 0407}, 058 (2004)
  [arXiv:hep-th/0405057].
}

\lref\morales{
  J.~F.~Morales and H.~Samtleben,
  ``Higher spin holography for SYM in d dimensions,''
  Phys.\ Lett.\ B {\bf 607}, 286 (2005)
  [arXiv:hep-th/0411246].
}

\lref\bianchitwo{
  N.~Beisert, M.~Bianchi, J.~F.~Morales and H.~Samtleben,
  ``On the spectrum of AdS/CFT beyond supergravity,''
  JHEP {\bf 0402}, 001 (2004)
  [arXiv:hep-th/0310292].
}

\lref\sphere{
  O.~Aharony, J.~Marsano, S.~Minwalla, K.~Papadodimas and M.~Van Raamsdonk,
  ``The Hagedorn / deconfinement phase transition in weakly coupled large N
  gauge theories,''
  Adv.\ Theor.\ Math.\ Phys.\  {\bf 8}, 603 (2004)
  [arXiv:hep-th/0310285].
}

\lref\rom{
  C.~Romelsberger,
  ``An index to count chiral primaries in N=1 d=4 superconformal field
  theories,''
  arXiv:hep-th/0510060.
}

\lref\mark{
  K.~Dasgupta, M.~M.~Sheikh-Jabbari and M.~Van Raamsdonk,
  ``Protected multiplets of M-theory on a plane wave,''
  JHEP {\bf 0209}, 021 (2002)
  [arXiv:hep-th/0207050].
}
\lref\kim{
  N.~Kim and J.~H.~Park,
  ``Superalgebra for M-theory on a pp-wave,''
  Phys.\ Rev.\ D {\bf 66}, 106007 (2002)
  [arXiv:hep-th/0207061].
}

\lref\cachazo{
 F.~Cachazo, M.~R.~Douglas, N.~Seiberg and E.~Witten,
  ``Chiral rings and anomalies in supersymmetric gauge theory,''
  JHEP {\bf 0212}, 071 (2002)
  [arXiv:hep-th/0211170].
}

\lref\hailin{
 H.~Lin and J.~Maldacena,
  ``Fivebranes from gauge theory,''
  arXiv:hep-th/0509235.
}

\lref\lubos{
  L.~Motl, A.~Neitzke and M.~M.~Sheikh-Jabbari,
  ``Heterotic plane wave matrix models and giant gluons,''
  JHEP {\bf 0306}, 058 (2003)
  [arXiv:hep-th/0306051].
}

\lref\wittenindex{
  E.~Witten,
  ``Constraints On Supersymmetry Breaking,''
  Nucl.\ Phys.\ B {\bf 202}, 253 (1982).
}

\lref\igor{
 S.~S.~Gubser, I.~R.~Klebanov and A.~W.~Peet,
  ``Entropy and Temperature of Black 3-Branes,''
  Phys.\ Rev.\ D {\bf 54}, 3915 (1996)
  [arXiv:hep-th/9602135].
}

\lref\reallone{
  J.~B.~Gutowski and H.~S.~Reall,
  ``Supersymmetric AdS(5) black holes,''
  JHEP {\bf 0402}, 006 (2004)
  [arXiv:hep-th/0401042].
}

\lref\reallthree{
  J.~B.~Gutowski and H.~S.~Reall,
  ``General supersymmetric AdS(5) black holes,''
  JHEP {\bf 0404}, 048 (2004)
  [arXiv:hep-th/0401129].
}

\lref\cvetic{
 Z.~W.~Chong, M.~Cvetic, H.~Lu and C.~N.~Pope,
  ``General non-extremal rotating black holes in minimal five-dimensional
  gauged supergravity,''
  arXiv:hep-th/0506029.
}

\lref\MinwallaKA{
S.~Minwalla,
``Restrictions imposed by superconformal invariance on quantum field
theories,''
Adv.\ Theor.\ Math.\ Phys.\  {\bf 2}, 781 (1998)
[arXiv:hep-th/9712074].
}

\lref\DolanZH{
F.~A.~Dolan and H.~Osborn,
``On short and semi-short representations for four dimensional superconformal
symmetry,''
Annals Phys.\  {\bf 307}, 41 (2003)
[arXiv:hep-th/0209056].
}

\lref\ascv{  A.~Strominger and C.~Vafa,
  ``Microscopic Origin of the Bekenstein-Hawking Entropy,''
  Phys.\ Lett.\ B {\bf 379}, 99 (1996)
  [arXiv:hep-th/9601029].
}

\lref\jmth{  J.~M.~Maldacena,
  ``Black holes in string theory,''
  arXiv:hep-th/9607235.
}

\lref\BeisertDI{
N.~Beisert, M.~Bianchi, J.~F.~Morales and H.~Samtleben,
``Higher spin symmetry and N = 4 SYM,''
JHEP {\bf 0407}, 058 (2004)
[arXiv:hep-th/0405057].
}

\lref\BianchiWX{
M.~Bianchi, J.~F.~Morales and H.~Samtleben,
``On stringy AdS(5) x S**5 and higher spin holography,''
JHEP {\bf 0307}, 062 (2003)
[arXiv:hep-th/0305052].
}

\lref\BianchiXI{
M.~Bianchi,
``Higher spins and stringy AdS(5) x S(5),''
arXiv:hep-th/0409304.
}

\lref\BianchiWW{
M.~Bianchi,
``Higher spin symmetry (breaking) in N = 4 SYM and holography,''
arXiv:hep-th/0409292.
}

\lref\BeisertTE{
N.~Beisert, M.~Bianchi, J.~F.~Morales and H.~Samtleben,
``On the spectrum of AdS/CFT beyond supergravity,''
JHEP {\bf 0402}, 001 (2004)
[arXiv:hep-th/0310292].
}

\lref\DobrevQV{
V.~K.~Dobrev and V.~B.~Petkova,
``All Positive Energy Unitary Irreducible Representations Of Extended
Conformal Supersymmetry,''
Phys.\ Lett.\ B {\bf 162}, 127 (1985).
}

\lref\DobrevVH{
  V.~K.~Dobrev and V.~B.~Petkova,
  ``On The Group Theoretical Approach To Extended Conformal Supersymmetry:
  Classification Of Multiplets,''
  Lett.\ Math.\ Phys.\  {\bf 9}, 287 (1985).
}

\lref\DobrevQZ{
  V.~K.~Dobrev and V.~B.~Petkova,
  ``Group Theoretical Approach To Extended Conformal Supersymmetry: Function
  Space Realizations And Invariant Differential Operators,''
  Fortsch.\ Phys.\  {\bf 35}, 537 (1987).
}

\lref\DobrevSP{
V.~K.~Dobrev and V.~B.~Petkova,
``All positive energy unitary
irreducible representations of the extended conformal superalgebra,''
Talk at the Symposium on Conformal Groups and Structures (Clausthal,
1985);
Proceedings, eds. A.O. Barut and H.D. Doebner, Lecture Notes in
Physics, Vol. 261 (Springer-Verlag, Berlin, 1986) pp. 300-308.
}

\lref\DobrevTK{
  V.~K.~Dobrev,
 ``Characters of the positive energy UIRs of D = 4 conformal supersymmetry,''
  arXiv:hep-th/0406154.
}

\lref\MackJE
\lref\MackJE{
G.~Mack,
``All Unitary Ray Representations Of The Conformal Group SU(2,2) With Positive
Energy,''
Commun.\ Math.\ Phys.\  {\bf 55}, 1 (1977).
}

\lref\BarsEP{
I.~Bars and M.~Gunaydin,
``Unitary Representations Of Noncompact Supergroups,''
Commun.\ Math.\ Phys.\  {\bf 91}, 31 (1983).
}

\lref\gunmar{
  M.~Gunaydin and N.~Marcus,
  ``The Spectrum Of The S**5 Compactification Of The Chiral N=2, D = 10
  Supergravity And The Unitary Supermultiplets Of U(2, 2/4),''
  Class.\ Quant.\ Grav.\  {\bf 2}, L11 (1985).
}

\lref\BerensteinJQ{
D.~Berenstein, J.~M.~Maldacena and H.~Nastase,
``Strings in flat space and pp waves from N = 4 super Yang Mills,''
JHEP {\bf 0204}, 013 (2002)
[arXiv:hep-th/0202021].
}

\lref\BeisertFV{
N.~Beisert, G.~Ferretti, R.~Heise and K.~Zarembo,
``One-loop QCD spin chain and its spectrum,''
arXiv:hep-th/0412029.
}

\lref\BeisertRY{
N.~Beisert,
``The dilatation operator of N = 4 super Yang-Mills theory and
integrability,''
Phys.\ Rept.\  {\bf 405}, 1 (2005)
[arXiv:hep-th/0407277].
}
\lref\Kac{
  V.~G.~Kac,
  ``Lie Superalgebras,''
  Adv.\ Math.\  {\bf 26}, 8 (1977).
}
\lref\BarabanschikovRI{
  A.~Barabanschikov, L.~Grant, L.~L.~Huang and S.~Raju,
  ``The spectrum of Yang Mills on a sphere,''
  arXiv:hep-th/0501063.
}

\lref\FuchsJV{
  J.~Fuchs and C.~Schweigert,
  ``Symmetries, Lie algebras and representations: A graduate course for
  physicists,'', Cambridge University Press (1997).
}

\lref\arfken{
G.~B.~Arfken and H.~J.~Weber,
``Mathematical methods for physicists'', Academic Press(2001).}

\lref\hong{ H.~Liu, ``Fine structure of Hagedorn transitions,''
arXiv:hep-th/0408001.
}

\lref\tloop{ O.~Aharony, J.~Marsano, S.~Minwalla, K.~Papadodimas and
M.~Van Raamsdonk, ``A first order deconfinement transition in large
N Yang-Mills theory on a
small S**3,''
Phys.\ Rev.\ D {\bf 71}, 125018 (2005) [arXiv:hep-th/0502149].
}

\lref\gl{ O.~Aharony, J.~Marsano, S.~Minwalla and T.~Wiseman,
``Black hole - black string phase transitions in thermal 1+1
dimensional supersymmetric Yang-Mills theory on a circle,'' Class.\
Quant.\ Grav.\  {\bf 21}, 5169 (2004) [arXiv:hep-th/0406210].
}

\lref\spenta{
L.~Alvarez-Gaume, C.~Gomez, H.~Liu and S.~Wadia,
``Finite temperature effective action, AdS(5) black holes, and 1/N
expansion,'' Phys.\ Rev.\ D {\bf 71}, 124023 (2005)
[arXiv:hep-th/0502227].
}

\lref\pallav{ P.~Basu and S.~R.~Wadia, ``R-charged AdS(5) black
holes and large N unitary matrix models,'' arXiv:hep-th/0506203.
}
\lref\oferun{
O. Aharony, unpublished.
}
\lref\BerensteinAA{
  D.~Berenstein,
  ``Large N BPS states and emergent quantum gravity,''
  arXiv:hep-th/0507203.
}

\lref\torus{ O.~Aharony, J.~Marsano, S.~Minwalla, K.~Papadodimas,
M.~Van Raamsdonk and T.~Wiseman, ``The phase structure of low
dimensional large N gauge theories on tori,'' arXiv:hep-th/0508077.
}

\lref\BerensteinKE{
  D.~Berenstein, C.~P.~Herzog and I.~R.~Klebanov,
  ``Baryon spectra and AdS/CFT correspondence,''
  JHEP {\bf 0206}, 047 (2002)
  [arXiv:hep-th/0202150].
}
\lref\BarnesBW{
  E.~Barnes, E.~Gorbatov, K.~Intriligator and J.~Wright,
  ``Current correlators and AdS/CFT geometry,''
  arXiv:hep-th/0507146.
}

\lref\bianchimixing{
  M.~Bianchi, B.~Eden, G.~Rossi and Y.~S.~Stanev,
  ``On operator mixing in N = 4 SYM,''
  Nucl.\ Phys.\ B {\bf 646}, 69 (2002)
  [arXiv:hep-th/0205321].  M.~Bianchi, G.~Rossi and Y.~S.~Stanev,
  ``Surprises from the resolution of operator mixing in N = 4 SYM,''
  Nucl.\ Phys.\ B {\bf 685}, 65 (2004)
  [arXiv:hep-th/0312228].
}

\lref\petkou{
  B.~Eden, A.~C.~Petkou, C.~Schubert and E.~Sokatchev,
  ``Partial non-renormalisation of the stress-tensor four-point function in  N
  = 4 SYM and AdS/CFT,''
  Nucl.\ Phys.\ B {\bf 607}, 191 (2001)
  [arXiv:hep-th/0009106].
}

\lref\bianchione{
  M.~Bianchi, J.~F.~Morales and H.~Samtleben,
  ``On stringy AdS(5) x S**5 and higher spin holography,''
  JHEP {\bf 0307}, 062 (2003)
  [arXiv:hep-th/0305052].
}

\lref\bianchitwo{
  N.~Beisert, M.~Bianchi, J.~F.~Morales and H.~Samtleben,
  ``On the spectrum of AdS/CFT beyond supergravity,''
  JHEP {\bf 0402}, 001 (2004)
  [arXiv:hep-th/0310292].
}

\lref\bianchithree{
  N.~Beisert, M.~Bianchi, J.~F.~Morales and H.~Samtleben,
  ``Higher spin symmetry and N = 4 SYM,''
  JHEP {\bf 0407}, 058 (2004)
  [arXiv:hep-th/0405057].
}

\lref\rossi{
  M.~Bianchi, S.~Kovacs, G.~Rossi and Y.~S.~Stanev,
  ``Properties of the Konishi multiplet in N = 4 SYM theory,''
  JHEP {\bf 0105}, 042 (2001)
  [arXiv:hep-th/0104016].
}
\lref\frolovone{
  G.~Arutyunov, S.~Frolov and A.~Petkou,
  ``Perturbative and instanton corrections to the OPE of CPOs in N = 4
  SYM(4),''
  Nucl.\ Phys.\ B {\bf 602}, 238 (2001)
  [Erratum-ibid.\ B {\bf 609}, 540 (2001)]
  [arXiv:hep-th/0010137].
}
\lref\frolovtwo{
  G.~Arutyunov, S.~Frolov and A.~C.~Petkou,
  ``Operator product expansion of the lowest weight CPOs in N = 4  SYM(4) at
  strong coupling,''
  Nucl.\ Phys.\ B {\bf 586}, 547 (2000)
  [Erratum-ibid.\ B {\bf 609}, 539 (2001)]
  [arXiv:hep-th/0005182].
}

\lref\edennew{
  B.~Eden, C.~Jarczak, E.~Sokatchev and Y.~S.~Stanev,
  ``Operator mixing in N = 4 SYM: The Konishi anomaly revisited,''
  Nucl.\ Phys.\ B {\bf 722}, 119 (2005)
  [arXiv:hep-th/0501077].
}

\lref\eden{
  B.~Eden, A.~C.~Petkou, C.~Schubert and E.~Sokatchev,
  ``Partial non-renormalisation of the stress-tensor four-point function in  N
  = 4 SYM and AdS/CFT,''
  Nucl.\ Phys.\ B {\bf 607}, 191 (2001)
  [arXiv:hep-th/0009106].
}

\lref\excep{
  G.~Arutyunov, B.~Eden, A.~C.~Petkou and E.~Sokatchev,
  ``Exceptional non-renormalization properties and OPE analysis of chiral
  four-point functions in N = 4 SYM(4),''
  Nucl.\ Phys.\ B {\bf 620}, 380 (2002)
  [arXiv:hep-th/0103230].
}

\vskip 0.8cm

\Title{\vbox{\baselineskip12pt\hbox{hep-th/0510251}}}
{\vbox{\centerline{An Index for 4 dimensional Super Conformal
Theories}}}

 \centerline{Justin Kinney$^{a}$, Juan Maldacena$^{b}$, Shiraz
Minwalla$^{c,d}$ and Suvrat Raju$^{d}$}
\smallskip

\centerline{\sl $^{a}$ Department of Physics, Princeton University,
Princeton, NJ 08544, USA} \centerline{\sl $^{b}$ Institute for
Advanced Study, Princeton, NJ 08540, USA} \centerline{\sl $^{c}$
Tata Institute of Fundamental Research, Mumbai 400005, India }
\centerline{\sl $^{d}$ Jefferson Physical Laboratory,  Harvard
University, Cambridge, MA 02138, USA }

\vskip .6in \centerline{\bf Abstract}{We present a trace formula for
an index over the spectrum of four dimensional superconformal field
theories on $S^3 \times $ time. Our index receives contributions
from states invariant under at least one supercharge and captures
all information -- that may be obtained purely from group theory --
about protected short representations in 4 dimensional
superconformal field theories. In the case of the $\CN=4$ theory our
index is a function of four continuous variables. We compute it at
weak coupling using gauge theory and at strong coupling by summing
over the spectrum of free massless particles in $AdS_5\times S^5$
and find perfect agreement at large $N$ and small charges. Our index
does not reproduce the entropy of supersymmetric black holes in
$AdS_5$, but this is not a contradiction, as it differs
qualitatively from the partition function over supersymmetric states
of the ${\cal N}=4$ theory. We note that entropy for some small
supersymmetric $AdS_5$ black holes may be reproduced via a D-brane
counting involving giant gravitons. For big black holes we find a
qualitative (but not exact) agreement with the naive counting of BPS
states in the free Yang Mills theory. In this paper we  also
evaluate and study the partition function over the chiral ring in
the $\CN=4$ Yang Mills theory. }

\smallskip
\Date{} 
\listtoc\writetoc

\newsec{Introduction}
Supersymmetry is a powerful tool for extracting exact information
about quantum field theories. Supersymmetry algebras that contain
R-charges in the right hand side have special BPS multiplets. These
multiplets occur at special values of energies or conformal
dimensions determined by their charge, and have fewer states than
the generic representation. An infinitesimal change in the energy of
a special multiplet turns it into a generic multiplet with a
discontinuously larger number of states. One might be tempted to use
this observation to conclude that the number of short
representations cannot change under variation of any continuous
parameter of the field theory; however there is a caveat. It is
sometimes possible for two or more BPS representations to combine
into a generic representation. For this reason only states that
cannot combine with other multiplets to  form a long representation
are guaranteed to be protected.

In this paper we construct some quantities, called indices, that
receive contributions only from those BPS states that cannot combine
into long representations. The indices that  we construct are
defined for 4 dimensional superconformal field theories (with
arbitrary number of supersymmetries) on $S^3 \times $ time. They
take the form \eqn\indform{{\cal I}^W=Tr[(-1)^F e^{\mu_i q_i}]} where $q_i$
are charges that commute with a particular supercharge. Our indices
closely resemble the Witten index \wittenindex, and are invariant
under all continuous deformations of the theory that preserve
superconformal invariance\foot{More generally they invariant under
all deformations of the theory that preserve the corresponding
supercharge.}. We demonstrate that our indices ${\cal I}^W$ contain all the
information about protected states that can be obtained by group
theory alone, and so should be useful in the study of general
super-conformal field theories.

The indices ${\cal I}^W$  are a functions of $2, 3 $ and $4$ continuous
variables for ${\cal N}=1,2, 4$ superconformal field theories
respectively. In the case of the $\CN=4$ Yang Mills theory we
explicitly compute this index ${\cal I}^W_{YM}$ in the free limit. Upon
taking the large $N$ limit the index receives contributions only
from states with energies of order one at all  chemical potentials
of order one. In other words ${\cal I}^W_{YM}$ does not undergo the
deconfinement phase transition described in \refs{\sundborg,
\sphere}. Moreover we find that ${\cal I}^W_{YM}$ agrees perfectly with the
index evaluated over the spectrum of free ten dimensional massless
fields propagating on $AdS_5 \times S^5$. This agreement provides a
check on the AdS/CFT conjecture in the BPS sector, which ends up
containing the same information as the matching of chiral primary
operators in \refs{\wittenhol,\oferun}.

Related to the fact that the index never `deconfines', in the limit
of very small chemical potential, with charges growing like $q \sim
N^2$ we find that the index ${\cal I}^{W}_{YM}$ grows rather slowly. In
particular, it does not grow fast enough to account for the entropy
of the BPS black holes in $AdS_5 \times S^5$ found in
\refs{\reallone,\reallthree,\cvetic}. This is not a contradiction
with $AdS/CFT$; the entropy of a black hole counts all
supersymmetric states with a positive sign whereas our index counts
the same states up to sign. It is possible for cancellations to
ensure that the Index is much smaller than the partition function
evaluated over supersymmetric states of the theory. This is
certainly what happens in the free $\CN=4$ theory, where both
quantities (the index and the partition function) may explicitly be
computed, and is presumably also the case at strong coupling.

It may well be possible to provide a weak coupling microscopic
counting of the entropy of BPS black holes
\refs{\reallone,\reallthree,\cvetic} in $AdS_5\times S^5$; however
such an accounting must incorporate some dynamical information about
${\cal N}=4 $ super Yang Mills beyond the information contained in
the superconformal algebra. In this paper we make some small steps
towards understanding the entropy of these black holes. In
particular we provide a counting of the entropy for small black
holes in terms of D-branes and giant gravitons in the interior. The
counting is rather similar to the one performed for the D1D5p black
holes \ascv . We also note that,  for large (large compared to the
AdS radius) black holes a naive computation of the simple partition
function of BPS states in the free theory gives a formula which has
similar features to the black hole answer.

The indices \indform\ do not exhaust all interesting calculable
information about supersymmetric states in all superconformal field
theory; in specific examples it is possible to extract more refined
information about supersymmetric states by adding extra input. An
explicit example where dynamical information allows us to make more
progress is the computation of the chiral ring
\refs{\oferun,\BerensteinAA}. In the case of $\CN=4$ Yang Mills theory, we write down explicit counting
formulas for 1/2, 1/4 and 1/8 BPS states. The counting can be done
in terms of $N$ particles in harmonic oscillator potentials. For
very large charges the entropy in these states grows linearly in
$N$.  By taking the large $N$ limit of these partition functions we
show that they display a second order phase transition which
corresponds to the formation of Bose-Einstein condensate.

The structure of this paper is as follows. In section 2 we review
the unitary representations of the conformal and superconformal
algebra, and list the linear combinations of (numbers of) short
representations that form indices in these algebras. In section 3 we
define the Witten Indices that are the main topic of this paper, and
explain how these are related to the indices of section 2. This
comparison allows us to argue that our Indices capture all group
theoretically protected information about short representations in
superconformal field theories. In section 4 we turn to the $\CN=4$
Super Yang Mills theory. We compute our index in this theory in free
Yang Mills on $S^3$ and show that it agrees perfectly, in the large
$N$ limit, with the same index computed over supergravitons in
$AdS_5\times S^5$. In section 5 we continue our study of
supersymmetric states in the $\CN=4$ Yang Mills theory on $S^3$. We
compute the partition function over ${1\over 16}^{th}$
supersymmetric states in free Yang Mills using perturbation theory
and in strongly coupled Yang Mills using gravity, and compare the two
results. In section 6 present exact formulas for the
partition function over  ${1 \over 4}^{th}$ and ${1\over 8}^{th}$
BPS states in $\CN=4$ Yang Mills. For large charges we find that the
free energy of these states scales linearly in $N$. This free energy
displays a second order transition which is associated to the
formation of a Bose-Einstein condensate.

This paper contains two related but distinct streams. Section 2 and
3 below concern themselves with the detailed nature of unitary
representations of the superconformal algebra. Sections 4, 5 and 6
study the supersymmetric states of the $\CN=4$ Yang Mills theory on
$S^3$. The link between these two streams is the Witten index,
defined in section 3.1.  The reader who is interested only in the
definition of the index and the results for ${\cal N}=4$ Yang Mills
could proceed directly to section 3.1 where the index is defined,
then to sections 4, 5  and 6 for computations in the $\CN=4$ Yang
Mills theory. On the other hand, the reader who is interested
principally in the algebraic aspects of this index, including the
demonstration that the Witten index captures all protected
information about superconformal field theories in four dimensions,
could focus on sections two and three.

In this revised version of the paper we have added section 6.2 which
shows, using the index, that a particular double trace operator in the
${\bf 20}$ of $SO(6)$ is protected.

While this paper was being completed we saw \rom\ which overlaps
with parts of section 3.

\newsec{4 dimensional Superconformal Algebras and their
Unitary Representations}
In this section we study the structure of representations of
conformal and superconformal algebras. Our goal is to understand
which representations, or combinations of representations, are
protected. This will allow us to show that all protected information
that can be obtained by using group theory alone is captured by the
index we will define in section 3.1. The reader who is willing to
accept this fact (and is otherwise uninterested in the structure of
unitary representations of the superconformal algebra), can just
jump to section 3.1 and from there to section 4. We start this
section with a discussion of the conformal algebra and then we
discuss the superconformal algebra.

\subsec{The 4 dimensional Conformal Algebra}
The set of killing vectors $M_{\mu \nu}= -i (x_\mu \p_\nu -x_\nu
\p_\mu)$, $P_{\mu}=- i \p_\mu$, $K_\mu= i (2 x_\mu x . \p -x^2
\p_\mu)$ and $H=x. \p$ form a basis for infinitesimal conformal
diffeomorphisms of $R^4$. These killing vectors generate the algebra
\eqn\conformalalg{\eqalign{[H,P_{\mu}] &= P_{\mu}, \cr [H,K_{\mu}]
&= - K_{\mu}, \cr [K_{\mu}, P_{\nu}] &= 2 (\delta_{\mu \nu} H - i
M_{\mu \nu}), \cr [M_{\mu \nu}, P_{\rho}] &= i(\delta_{\mu \rho}
P_{\nu} - \delta_{\nu \rho} P_{\mu}) ,\cr [M_{\mu \nu},K_{\rho}] &=
i (\delta_{\mu \rho} K_{\nu} - \delta_{\nu \rho} K_{\mu}), \cr
[M_{\mu \nu}, M_{\rho \sigma}] &= i \left(\delta_{\mu \rho} M_{\nu
\sigma} + \delta_{\nu \sigma} M_{\mu \rho} - \delta_{\mu \sigma}
M_{\nu \rho} - \delta_{\nu \rho} M_{\mu \sigma} \right) .\cr } }

Consider a 4 dimensional Euclidean quantum field theory. It is
sometimes possible to combine the conformal killing symmetries of
the previous paragraph with suitable action on fields to generate a
symmetry of the theory. In such cases the theory in question is
called a conformal field theory (CFT). The Euclidean path integral of
a CFT may be given a useful Hilbert space interpretation via a
radial quantization. Wave functions (kets) are identified with the
path integral, with appropriate operator insertions,  over the unit
3 ball surrounding the origin. Dual wave functions (bras) are
obtained by acting on kets with by the conformal symmetry
corresponding to inversions $x^\mu ={x^\mu \over x^2}$ \foot{As a
consequence, a bra may be thought of as being generated by a path
integral, performed with appropriate insertions, on $R^4$ minus the
unit 3 ball. The scalar product between a bra and a ket is the path
integral - with insertions both inside and outside the unit sphere -
over all of space. }. Under an inversion, the killing vectors of the
previous paragraph transform as $M_{\mu \nu} \rightarrow M_{\mu
\nu}$, $H \rightarrow -H$, $P_\mu \rightarrow K_\mu$. As a
consequence, the operators $M_{\mu \nu}, P_\mu, K_\nu$ are
represented on the CFT Hilbert space \conformalalg\  with the
hermiticity conditions \eqn\herm{M_{\mu \nu}= M_{\mu \nu}^\dagger,
~~~P_\mu = K_\mu^\dagger.} Radial quantization of the CFT on $R^4$
is equivalent to studying the field theory on $S^3 \times$ time. The
operators $M_{\mu \nu}$ generate the $SO(4)$ rotational symmetries
of $S^3$, and $H$ is the Hamiltonian. From this point of view the
conjugate generators $P_\mu$ and $K_\mu$ are less familiar; they act
as ladder operators, respectively raising and lowering energy by a
single unit.


The Hilbert space of a CFT on $S^3 \times $ time may be decomposed
into a sum of irreducible unitary representations of the conformal
group. The theory of these representations was studied in detail by
\MackJE. We present a brief review below, as a warm up for the
superconformal algebra (see \suvrat\ and references therein for a
recent discussion).

\subsec{Unitary Representations of the Conformal Group}

Any irreducible representation of the conformal group can be written
as some direct sum of representations, $R^i_{compact}$, of the
compact subgroup $SO(4)\times SO(2)$: \eqn\reduction{ R_{SO(4,2)}=
\sum_i \bigoplus R_{compact}^i. } The states within a given $SO(4)
\times SO(2)$ representation all have the same energy. As the energy
spectrum of any sensible quantum field theory is bounded from below,
the representations of interest to us all possess a particular set
of states with minimum energy. We will call these states (which we
will take to transform as $R_{compact}^{\lambda}$) the lowest weight
states. The $K^{\mu}$ operators necessarily annihilate all the
states in $R_{compact}^{\lambda} $ because the $K^{\mu}$ have
negative energy. We can now act on these lowest weight states with
an arbitrary number of $P_{\mu}$ (`raising') operators to generate
the remaining states in the representation. We will use the charges
of the lowest weight state $|\lambda \rangle \equiv |E,j_1,j_2
\rangle$ to label this representation. We use the fact that
$SO(4)=SU(2) \times SU(2)$; $j_1$ and $j_2$ are standard
representation labels of these $SU(2)s$.

It is important that not all values of $E, j_1, j_2$ yield unitary
representations of the conformal group. For a representation to be
unitary, it is necessary for all states to have positive norm.
Acting on the lowest energy states with $P_\mu$, we obtain (via a
Clebsh Gordan decomposition) states that transform in the
representations $(E+1,j_1 \pm {1/2}, j_2 \pm {1/2})$. The norm of
these states may be calculated using the commutation
relations \conformalalg\ \MinwallaKA . The states with lowest norm turn out to
have quantum numbers $(E+1, j_1 - \half, j_2-\half)$, and this norm
is given by \eqn\normc{ {\|~~\|^2  \over 2} =E-j_1-1 +\delta_{j_1 0} -j_2 -1
+\delta_{j_2 0}.}

Unitarity then requires that \eqn\unitarity{ \eqalign{ (i) &\quad E
\ge j_1 + j_2 +2 \quad j_1 \ne 0 \quad j_2 \ne 0, \cr (ii) &\quad E
\ge j_1 + j_2 + 1 \quad j_1 j_2 = 0. }}
 The special case
$j_1=j_2=0$ has an additional complication. In this case the norm of
the level 2 state $P^2 |\psi\rangle$ is proportional \MinwallaKA\ to
$E(E-1)$ and so is negative for $0<E<1$. The representation with
$E=0$ is annihilated by all momentum operators and represents the
vacuum state. The representation at $E=1$ is short and it obeys the
equation $P^2|E\rangle =0$ so it is a free field in the conformal
field theory.

Unitary representations exist even when this bound is {\it strictly
saturated}. The zero norm states, and all their descendants, are
simply set to zero in these representations \foot{The consistency of
this procedure relies on the fact that, at the unitarity bound,
zero norm states are orthogonal to all states in the representation. As a consequence
the inner product on  the representation modded out by zero norm states is well defined
and positive definite.} making them shorter than generic.

Now consider a one parameter (fixed line) of conformal field
theories. An infinitesimal variation of the parameter that labels the
theory will, generically, result in an infinitesimal variation in the
energy of all the long representations of the theory. However a
short representation can change its energy only if it turns into a
long representation. In order for this to happen without a
discontinuous jump in the spectrum of the CFT (i.e. a phase
transition), it must pair up with some other representation, to make
up the states of a long representation with energy at just above the
unitarity threshold. Groups of short representations that can pair
up in this manner are not protected; the numbers of such
representations can jump discontinuously under infinitesimal
variations of a theory.

However consider an index $I$ that is defined as a sum of the form
\eqn\indexdef{ I= \alpha[i] n[i]} where $i$ runs over the various
short representations of the theory, $n[i]s$ are the number of
representations of the $i^{th}$ variety, and $\alpha[i]$ are fixed
numbers chosen so that $I$ evaluates to zero on any collection of
short representations that can pair up into long representations. By
definition such an index is unaffected by groups of short
representations pairing up and leaving, as it evaluates to zero
anyway on any set of representations that can. It follows that an
index cannot change under continuous deformations of the theory.

We will now argue that the conformal algebra does not admit any non
trivial indices. In order to do this we first list how a long
representation decomposes into a sum of other representations (at
least one of which is short) when its energy is decreased so that it
hits the unitarity bound. Let us denote the representations as follows.
  $A_{E, j_1, j_2}$ denotes the generic long representation,
  $C_{j_1, j_2}$  denotes the short
representations  with $j_1$ and $j_2$ both not equal to zero,
 $B^L_{j_1}$ denotes   the short representations with $j_2=0$, $ B^R_{j_2}$
 the short ones with
  $j_1=0$. Finally we denote the
special short representation at $E=1$ and $j_1=j_2=0$ by $B$.
As the energy is decreased to approach the unitarity bound we find
\eqn\confchar{\eqalign{& \lim_{\epsilon
\to 0} \chi[A_{j_1+j_2+2 + \epsilon, j_1, j_2}]= \chi[ C_{j_1, j_2}]+ \chi[
A_{j_1+j_2+3, j_1 - \half, j_2 -\half}]  \cr &
\lim_{\epsilon \to 0}\chi[A_{j_1+1+ \epsilon, j_1,
0}]=\chi[B^L_{j_1}]+ \chi[C_{ j_1 - \half, \half}] \cr &
\lim_{\epsilon \to 0} \chi[A_{j_2+1+ \epsilon, 0,
j_2}]=\chi[B^R_{j_2}]+ \chi[C_{ \half, j_2 - \half}] \cr &
\lim_{\epsilon \to 0} \chi[A_{1+\epsilon, 0,0}]= \chi[B]+
\chi[A_{3,0,0}].}} In \confchar\ and throughout this paper, the
symbol $\chi$ denotes the super-character on a representation\foot{
i.e. $Tr_R (-1)^F G$ where $R$ is an arbitrary representation, $G$
is an arbitrary group element, and $F$ is the Fermion number, which
plays no role in the representation theory of the conformal group,
but will be important when we turn superconformal groups below.}.

It follows from \confchar\ that $\sum_i \alpha_i n_i$ is an index
only if \eqn\alp{\alpha_{C_{j_1,j_2}}=0, ~~~ \alpha_{B^L_{j_1}}+
\alpha_{C_{ j_1 - \half, \half}}=0, ~~~ \alpha_{B^R_{j_2}}+
\alpha_{C_{ \half, j_2 - \half}}=0, ~~~\alpha_{B}=0.} The only
solution to these equations has all $\alpha$ to zero; consequently
the conformal algebra admits no nontrivial indices. The
superconformal algebra will turn out to be more interesting in this
respect.

\subsec{Unitary Representations of $d=4$ Superconformal Algebras}
In the next two subsections we review the unitary representations of the $d=4$
 superconformal algebras \Kac\ that were
 studied in \refs{\DobrevQV,\DobrevVH, \DobrevSP,\DobrevQZ, \MinwallaKA,
\DolanZH, \DobrevTK}.
A supersymmetric field theory that is also conformally invariant,
actually enjoys superconformal symmetry, a symmetry algebra that is
larger than the union of conformal and super symmetry algebras. The
bosonic subalgebra of the $\CN=m$ superconformal algebra consists of
the conformal algebra times $U(m)$, except in the special case
$m=4$, where the $R$ symmetry algebra is $SU(4)$. The fermionic
generators of this algebra consist of  the $4m$ supersymmetry
generators $Q^{ \alpha i }$ and $\bar{Q}_{i}^{\dot \alpha}$,
together with the super conformal generators $S_{  \alpha i},
\bar{S}^{j}_{\dot{\alpha}}$. The generators transform under $SO(4)
\times U(m)$ as indicated by their index structure (an upper $i$
index indicates a $U(m)$ fundamental, while a lower $i$ index is a
$U(m)$ anti-fundamental). The commutation relations of the algebra
are listed in detail in Appendix A.1. In particular,
\eqn\QScommut{\{S_{\alpha i}, Q^{\beta j} \} = \delta^{j}_{i}
(J_1)^{\beta}_{\alpha} + \delta^{\beta}_{\alpha} R^{j}_{i} +
 \delta^{j}_{i} \delta^{\beta}_{\alpha} ({H \over 2} + r{4 - m \over 4 m})}
where $(J_1)^{\beta}_{\alpha}$ are the $SU(2)$
generators in spinor notation, $R^j_i$ are the
$SU(m)$ generators and $r$ is the $U(1)$ generator.
As in the previous subsection, radial quantization
endows these generators with hermiticity properties; specifically
\eqn\hermitm{(Q^{  \alpha i })^{\dagger} = S_{  \alpha i },
~~~(\bar{Q}^{ \dot{\alpha}}_{i})^\dagger=\bar{S}^{i}_{\dot \alpha} }

The theory of unitary representations of the superconformal algebra
is similar to that of the conformal algebra. Irreducible
representations are labeled by the energy $E$ and the  $SU(2) \times
SU(2)$ and $U(m)$ representations of their lowest weight states. We
label $U(m)$ representations by their $U(1)$ charge $r$ (normalized
such that $Q^{ \alpha i}$ has charge +1 and $\bar{Q}_{i}^{ {\dot
\alpha}}$ has charge -1) and  the integers $R_k$ $(k=1 \ldots m-1)$,
the number of columns of height  $k$ in the Young Tableaux for the
representation.\foot{$R_k$ may also be thought of as the eigenvalues
of the highest weight vectors under the diagonal generator $R_k$
whose $k^{th}$ diagonal entry is unity, $(k+1)^{th}$ entry is $-1$,
and all other are zero, in the defining representation.}

Lowest weight states are annihilated by the $S$ but, generically,
not by the $Q$ operators.  $Q^{\alpha i}$ have $E=\half$ and
transform in the $SU(2) \times SU(2)\times U(m)$ representation with
quantum numbers $ j_1=\half, ~~j_2=0, ~~r=1, ~~R_1=1, ~~R_i=0
~~(i>1).$ Let $|\psi_a \rangle$ be the set of lowest weight states
of this algebra that transforms in the representation $(E, j_1, j_2,
r, R_i)$. The states $Q^{\alpha i} |\psi_a\rangle $ transform in all
the Clebsh Gordan product representations; the lowest norm among
these states occurs for those that have quantum numbers $(E+\half,
j_1-\half, j_2, r+1, R_1+1, R_j)$; the value of the norm of these
states is given by \MinwallaKA\ \eqn\norm{\eqalign{2  \| \chi_1 \|^2
& = E+2 \delta_{j_1,0}-E_1(j_1, r, R_i) ~,  \cr E_1 &\equiv   2 +2
j_1 + 2 {\sum_{p=1}^{m-1} (m-p)R_p \over m} +{r(4-m) \over 2 m}.}}
In a similar fashion, of the states of the form $\bar{Q}_{  {\dot
\alpha} i} |\psi\rangle$ the lowest norm occurs for those that
transform in $(E+\half, j_1, j_2-\half, r-1, R_k, R_{m-1}+1)$, and
the norm of these states is equal to \MinwallaKA\
\eqn\normt{\eqalign{2 \| \chi_2 \|^2 &=E+2 \delta_{j_2,0} -E_2(j_1,
j_2, r, R_i) \cr E_2&\equiv  2 + 2 j_2 +
 {2 \sum_{p=1}^{m-1} p R_p  \over m} - {r (4-m) \over
2 m} .}}

Clearly unitarity demands that $\|\chi_1 \|^2 \geq 0$ and $\| \chi_2 \|^2 \geq 0$.
As for the conformal group, representations with either
$\|\chi_1 \|^2 =0$ or $\|\chi_2 \|^2 =0$ or both zero
are allowed. In such representations
the zero norm states and all their descendants  are simply set to
zero, yielding short representations.

In the special case $j_1=0$ the positivity of the norm at level 2
yields more information.  Of states of the form $Q^{\alpha i}
Q^{\beta j} |\psi_a\rangle$ (where $|\psi_a\rangle$ are the lowest
weight states), those that have the smallest norm transform in the
representation $(E+1, 0, j_2, r+2, R_1+2, R_j)$. The norm of these
states turns out to be proportional to $(E-E_1)(E-E_1 + 2)$ where
$E_1$ is defined in \norm. Thus unitarity disallows representations
in the window $E_1-2 <E<E_1$. Representations at $E=E_1-2$ and
$E=E_1$ are both short and both allowed. Representations at
$E=E_1-2$ are special because they are separated from long
representations (with the same value of all other charges) by an
energy gap of two units. All these remarks also apply to the special
case $j_2=0$, upon replacing $Q^{\alpha i}$ with $\bar{Q}^{{\dot
\alpha} }_i$ and $E_1$ with $E_2$.

In \DolanZH, Dolan and Osborn, performed a comprehensive tabulation of
 short representations of the $d=4$ superconformal algebras.
 We will adopt a notation that is slightly different from theirs.
 Representations are denoted by ${\bf x^L
x^R}_{E, j_1, j_2, r, R_i}$ where
\eqn\notation{{\bf x^L}=\cases{{\bf a} & if
$E>E_1$ \cr {\bf c} & if $E=E_1$ and $j_1 \geq 0$ \cr {\bf b} & if $E=E_1-2$ and $j_1=0$ }} and
\eqn\notation{{\bf x^R}=\cases{{\bf a} & if $E>E_2$ \cr {\bf c} & if $E=E_2$ and $j_2
\geq 0$
 \cr {\bf b} & if $E=E_2-2$ and $j_2=0$
}}
Further, we will usually omit to specify the first (energy)
subscript on all short representations as this energy is determined
by the other charges. Thus representations denoted by ${\bf aa}$ are long;
all other representations are short.

\subsec{The Null Vectors in Short Representations} We now study the
nature of the null vectors in short representations in more detail.
Consider a representation of the type ${\bf cx}$, with $j_1>0$,
where ${\bf x}$ is either ${\bf a}, {\bf c}$ or ${\bf b}$. Such a
representation has $\| \chi_1 \|^2=0$.  The descendants of the
null-state form another representation of the superconformal
algebra. This representation also has null states\foot{When we say
that a short representation has  `null states' of a particular type
we mean the following. When we lower the energy of a long
representation down to its unitarity bound ($E_1 $ or $E_2$), the
long representation splits into a positive norm short representation
$m$, plus a set of null representations $m'$. We describe this
situation by the words `the short representation $m$ has null
representations $m'$. As is clear from this definition, it is
meaningless to talk of the null state content of representations of
the sort ${\bf bx}$ or ${\bf xb}$, as these representations are
separated by a gap from long representations.} characterized by
their own value of $(\|\chi'_1\|^2, \|\chi'_2\|^2)$. A short
calculation \foot{$(\|\chi'_1\|^2, \|\chi_2'\|^2)=(\half +1
-2(m-1)/m-(4-m)/2m, \|\chi_2\|^2 - \half +2/m +(4-m)/2m)=(0,
\|\chi_2\|^2 )$.}shows that $ \|\chi'_1\|^2, \|\chi'_2\|^2)/ =(0,
\|\chi_2\|^2)$ . It follows that the $Q$ null states of a
representation of type ${\bf cx}$ are generically also of the type
${\bf cx}$. The exception to this rule occurs when $j_1=0$, in which
case the null states are of type ${\bf bx}$. Of course analogous
statements are also true for ${\bar Q}$ null states. All of this may
be summarized in a set of decomposition formulae, for the
supercharacters, \eqn\scdef{ \chi[R] ={\rm Tr}_{R} \left[ (-1)^{2
(J_1 + J_2)} G \right],} where $G$
 is an arbitrary element of the superconformal group. These formulae describe how a long
representation decomposes into a set of short representation  when
its energy hits the unitarity bound. \eqn\charac{\eqalign{ \lim_{\ep
\to 0} \chi[{\bf aa}_{E_1+ \ep, j_1, j_2, r, R_i}]&= \chi[{\bf \tilde{c}a}_{j_1,
j_2, r, R_i}]+\chi[{\bf \tilde{c}a}_{j_1-\half, j_2, r+1, R_1+1, R_j}],\, E_1>E_2
 \cr
\lim_{\ep \to 0} \chi[{\bf aa}_{E_2+ \ep, j_1, j_2, r, R_i}]& =
\chi[{\bf a\tilde{c}}_{j_1, j_2, r, R_i}]+ \chi[{\bf a\tilde{c}}_{j_1, j_2
-\half, r-1, R_k, R_{m-1}+1} ],\, E_2 > E_1 \cr \lim_{\ep \to 0}
\chi[{\bf aa}_{E_2+ \ep, j_1, j_2, r, R_i}]& =
\chi[{\bf \tilde{c}\tilde{c}}_{j_1, j_2, r, R_i}]+
\chi[{\bf \tilde{c}\tilde{c}}_{j_1 -\half, j_2, r+1, R_1+1, R_j}]+ \cr
&\chi[{\bf \tilde{c}\tilde{c}}_{j_1, j_2 -\half, r-1, R_k, R_{m-1}+1} ] +
\chi[{\bf \tilde{c}\tilde{c}}_{j_1-\half, j_2 -\half, r, R_{1}+1, R_l,
R_{m-1}+1}], \, E_1=E_2 }} where, in this equation and, as far as
possible, in the rest of the paper, we use the index convention
\eqn\indexconv{i = 1 \ldots m-1, ~~~ j = 2 \ldots m-1, ~~~k = 1
\ldots m-2, ~~~l = 2 \ldots m-2.} On the right hand side  of
\charac\ we have used the notation given in table 1.
\bigskip
{\hsize=2.3in\parindent=0pt
\leftline{\bf Table 1 : Short Representations}
\nobreak
\vskip 3pt
\valign{\hrule # \strut \hrule &# \strut &# \strut  \hrule&#\strut&#\strut \hrule&#\strut&#\strut&#\strut&#\strut\hrule \cr
\noalign{\vrule}
\hfil {\bf Symbol} \hfil &\multispan2 \vfil
\hfil ${\bf {\tilde c}a}_{j_1, j_2, r, R_i}$ \vfil \hrule &\multispan 2
 \vfil \hfil ${\bf a{\tilde c}}_{j_1, j_2, r, R_i}$ \vfil  \vfil \hrule
  &\multispan4 \vfil  \hfil ${\bf {\tilde c} {\tilde c}}_{j_1, j_2, r, R_i}$ \vfil \hrule
  \cr \noalign{\vrule}
\hfil {\bf Denotation} \hfil
&\hfil $ {\bf ca}_{j_1, j_2, r, R_i}
 ,~~j_1 \geq  0 $
 & \hfil $
{\bf ba}_{0, j_2, r+1, R_1+1, R_j} , ~~j_1={-\half} $&\hfil $
{\bf ac}_{j_1, j_2, r, R_i}
 ,~~j_2 \geq  0 $&
\hfil $
{\bf ab}_{j_1, 0, r-1, R_k, R_{m-1}+1} , ~~j_2={-\half} $&\hfil $
{\bf cc}_{j_1, j_2, r, R_i}
 ,~~ j_1 \geq  0, j_2 \geq 0 $&
\hfil $
{\bf cb}_{j_1,0, r-1, R_k,R_{m-1} + 1} , ~~j_2 = {-1 \over 2}, j_1 \geq 0  $&\hfil $
{\bf bc}_{0,j_2, r+1, R_1+1,R_j} ,~~j_1 = {-\half}, j_2 \geq 0 $&\hfil $
{\bf bb}_{0,0,r, R_1+1,R_l, R_{m-1} + 1} ,~~j_1 = j_2 ={-\half}  $ \hfil
\cr \noalign{\vrule}
}}

\subsec{Indices For Four Dimensional Super Conformal Algebras}
We now turn to a study of the indices for these algebras. Using \charac\
it is not difficult to convince oneself that the set of Indices for the
four dimensional superconformal field theories is a vector space that is spanned by
\item{1.} The number of representations of the sort ${\bf bx}$ with
$R_1=0$ or $R_1=1$ plus the number of representations of the sort
${\bf xb}$ with $R_{m-1}=0$ or $R_{m-1}=1$.
\item{2.} The Indices
\eqn\firstind{I^L_{j_2, \hat r, M, R_j} =
\sum_{p=-1}^{M} (-1)^{p+1} n[{\bf \tilde{c}x}_{{p\over 2}, j_2, \hat r-p, M-p,
R_j}]} for all values of $\hat r$ and non negative integral values of
$j_2, M, R_j$. In the case $m=1$ we do not  have the indices $M$ or $R_j$
 and the sum runs from
$p=-1$ to infinity. In the $m=4$ case, simply ignore the $r$ and $\hat r$ subindices.
\item{3.} The Indices
\eqn\secondind{I^R_{j_1, r'',R_k, N} =
\sum_{p=-1}^{M} (-1)^{p+1} n[{\bf x\tilde{c}}_{j_1, {p\over 2}, r''+p, R_k,
N-p}]} for all values of $r''$ and non negative integral values of
of $j_1, R_k, N$, with the same remarks for $m=1,4$.

In the special case that representations that contribute to the sum
in \firstind\ and \secondind\ have quantum numbers on which
$E_1=E_2$\foot{If this relation is true for any term that
contributes to the sum, it is automatically true on all other terms
as well.}, the indices \firstind\ and \secondind\ are subject to the
additional constraints \eqn\consttrr{\sum_{p=-1}^N (-1)^p
I^L_{{p\over 2}, r'''+p, M, R_l, N-p}= \sum_{p=-1}^M (-1)^p
I^R_{{p\over 2}, r'''-p, M-p, R_l, N} ~~~(E_1=E_2)} for all values
of $r'''= -\infty \ldots \infty$, and non negative integral values
of $M, N, R_l$.

These results are explained in more detail in Appendix B.1, where we
also present a detailed
listing of the absolutely protected multiplets, for the $\CN=1, 2,
4$ superconformal algebras.

\newsec{A Trace Formula for the Indices of Superconformal Algebra}
The supercharges $Q^{\alpha i } $ transform in the fundamental or
$(1, 0, \ldots ,0)$ representation of $SU(m)$. Let $Q \equiv Q^{
{-\half},1}$, the supercharge whose $SU(2) \times SU(2)$ quantum
numbers are $(j_1^3,j_2^3) = (-\half, 0)$, that has $r=1$, and whose
$SU(m)$ quantum numbers are $(1, 0, \ldots 0)$. Let $S \equiv
Q^\dagger$. Then (see \QScommut) \eqn\qscom{ 2 \{S, Q \}=
H-2J_1-2\sum_{k=1}^{m-1} {m-k \over m}R_k-{(4-m)r \over 2m} = E-
(E_1-2)  \equiv \Delta.} It follows from \qscom\ that every state in
a unitary representation of the superconformal group has $\Delta
\geq 0$. Note that the Jacobi identity implies that $Q$ and $S$
commute with $\Delta$.

Consider a unitary representation $R$ of the superconformal group
that is not necessarily irreducible. Let $R_{\Delta_0}$ denote the
linear vector space of states with $\Delta = \Delta_0 >0$. It
follows from \qscom\ that if $|\psi \rangle$ is in $R_{\Delta_0}$
then \eqn\sid{|\psi\rangle = Q {S \over \Delta_0}| \psi \rangle + S{
Q \over \Delta_0} | \psi \rangle.} Let $R^Q_{\Delta_0}$ denote the
subspace of $R_{\Delta_0}$ consisting of states annihilated by $Q$
and $R^S_{\Delta_0}$ the set of states in $R_{\Delta_0}$ that are
annihilated by $S$. It follows immediately from \sid\ (and the
unitarity of the representation) that
$R_{\Delta_0}=R^Q_{\Delta_0}+R^S_{\Delta_0}$ and that $S
|\psi\rangle = |\psi'\rangle$ is a one to one map from $R_{\Delta_0}^Q
$ to $R_{\Delta_0}^S$ ($Q/\Delta_0$ provides the inverse map).

Now consider the Witten index
\eqn\winddef{{\cal I}^{WL}=Tr_R \left[ (-1)^F \exp(-\beta
\Delta + M) \right]}  where
$M$ is any element of the subalgebra of the superconformal algebra
that commutes with $Q$ and $S$. We discuss this subalgebra in detail
in the next subsection.   It follows that the states in
$R_{\Delta_0}$ do not contribute to this index, the contribution of
$R_{\Delta_0}^Q$ cancels against that of $R_{\Delta_0}^S$.
Consequently, ${\cal I}^{WL}$ receives contributions only from states with
$\Delta=0$, i.e. those states that are annihilated by both $Q$ and
$S$. Thus,   despite appearances,  \winddef\ is independent of
$\beta$.  As no long representation contains states with $\Delta=0$,
such representations do not contribute to ${\cal I}^{WL}$. It also follows
from continuity that ${\cal I}^{WL}$ evaluates to zero on groups of short
representations that a long representation breaks up into when it
hits unitarity threshold. As a consequence ${\cal I}^{WL}$ is an index; it
cannot change under continuous variations of the superconformal
theory, and must depend linearly on the Indices, $I^L$ and $I^R$,
listed in the previous section. We will explain the relationship
between ${\cal I}^{WL}$ and $I^{L}$ in more detail in subsection 3.2 and
3.3 below. The main result of the following subsections is to show
that \winddef\ (and its ${\cal I}^{WR}$ version) completely capture  the
information contained in the indices defined in the previous
section, which is all the information about protected
representations that can be obtained without invoking any dynamical
assumption. In appendix B.2 we derive most of the results of section
2 in a way that uses a smaller amount of group theory.

\subsec{The Commuting Subalgebra}
In this subsection we briefly describe the
subalgebra of the superconformal algebra that commutes with the
$SU(1 |1)$ algebra spanned by $Q, S, \Delta$.

The $\CN=m, d=4$ superconformal algebra is the super-matrix algebra
$SU(2, 2|m)$\foot{For $m=4$ we have $PSU(2,2|4)$.}. Supersymmetry
generators transform as bifundamentals under the bosonic subgroups
$SU(2,2)$ and $SU(m)$. It is not difficult to convince oneself that
the commuting  subalgebra we are interested in is
$SU(2,1|m-1)$\foot{ Or $PSU(2,1|3)$ for $m=4$.}. The generators of
 $SU(2,1 |m-1)$ are related to those of $SU(2,2 |m)$ via the
obvious reduction. In more detail, the bosonic subgroup of
$SU(2,1|m-1)$ is $SU(2,1)\times U(m-1)$. The $U(m-1)$ factor sits
inside the superconformal $U(m)$ setting all elements the first row
and first column to zero, except for the 11 element which is set to
one. The Cartan elements $(E',j_2',r',R_i')$ of the subalgebra are
given in terms of those for the full algebra by
\eqn\qnfirst{E'=E+j_1, ~~~j'_2=j_2,~~~ r' ={(m-1) r \over m}
-\sum_{p=1}^{m-1}{m-p \over m} R_p, ~~~R_k'=R_{k+1}.} where $R'_k$
are the Cartan elements of $U(m-1)$ and $r'$ is the $U(1)$ charge in
$U(m-1)$. We will think of \qnfirst\ as defining a (many to one) map
from $(E,j_1,j_2,r,R_i)$ to $(E',j_2',r',R'_i)$

We will be interested in the representations of the subalgebra,
$SU(2,1|m-1)$, that are obtained by restricting a representation of
the full algebra, $SU(2,2|m)$, to states with $\Delta = 0$. Null
vectors, if any, of $SU(2,1|m-1)$ are inherited from those of
$SU(2,2|m)$. Is is possible to show
  that $SU(2,1|m-1)$ is short only when $SU(2,2|m)$ is one of the
representations $ c b, c c$ or if $R$ is $b x$ with $R_1 = 0$.
See  Appendix B.3 for further discussion.

\subsec{${\cal I}^{WL}$ expanded in sub-algebra characters with $I^L$ as
coefficients} In this subsection we present a formula for Index
${\cal I}^{WL}$ as a sum over super characters of the commuting subalgebra,
$SU(2,1|m-1)$, more details may be found in Appendix B.2.

It is not difficult to convince oneself (see Appendix B.2) that on
any short irreducible representation $R$ of the superconformal
algebra $SU(2,2 | m)$, ${\cal I}^{WL}$ evaluates  to the supercharacter of a
single irreducible representation $R'$ of the subalgebra
$SU(2,1 | m-1)$. More specifically we find
\eqn\indexchar{ \eqalign{
{\cal I}^{WL}[{\bf bx}_{0,j_2,r,R_i}] =& \chi_{sub}[\vec b]
\cr
{\cal I}^{WL}[{\bf cx}_{j_1, j_2, r,
R_i}]=& (-1)^{2 j_1+1} \chi_{sub}[{\vec c}]
}}
where  $\chi_{sub} $ is the
supercharacter \eqn\schartwo{\chi_s[R']={\rm Tr}_{R'}
\left[ (-1)^{2 J_2} G'\right] ,}
 where $G'$ is an element of the Cartan subgroup. The vectors $\vec{b}$ and $\vec{c}$ specify the highest weight of the representation of the subalgebra in the Cartan basis $[E',j_2',r',R'_{k}]$ defined in \qnfirst.
\eqn\defqbc{ \eqalign{
 \vec{b} & = [ {3\over 2}r-2r',j_2,r',R_j], \cr
 \vec{c} & =[ 3 +{3} (j_1 + r/2)-2r',j_2,r',R_j]
 }}
where  $r'$ is the function defined in
\qnfirst; we emphasize the fact that it depends on $r$ and $R_1$ only
through the combination $r-R_1$.

Notice that the functions that specify the character of the subalgebra, \defqbc ,
are not one to one.
 In fact, it follows from \indexchar, \defqbc,  that ${\cal I}^{WL}$ evaluates to
the same subalgebra character for each representation $R$ that
appears in the sum in \firstind, for fixed values of $j_2, r', M,
R_i$. Notice that by formally setting $j_1=-1/2$ in the second line
of \defqbc\ we get the Cartan values for the subalgebra that we
expect for the representation ${\bf b}$ according to the definition
of ${\bf \tilde c}$ in table 1. This implies that we can replace
${\bf c}$ in \indexchar\ by ${\bf \tilde c}$.
 More specifically
\eqn\indconc{
{\cal I}^{WL}[{\bf \tilde c}_{{p \over 2} , j_2, {\hat r}- p, M-p,R_j} ] = (-1)^{p} {\cal I}^{WL}[
{\bf \tilde c}_{0,j_2,{\hat r}, M, R_j} ]
}
It follows immediately from \indexchar, \indconc, that ${\cal I}^{WL}$, evaluated on
any (in general reducible representation) $A$ of the superconformal algebra evaluates to
\eqn\fineq{ \eqalign{ {\cal I}^{WL}[A]=&\sum_{j_2, r, R_i} \left( n[{\bf bx}_{0, j_2, r, 0, R_i}]
\chi_{sub}[{{\vec b_0}}] + n[{\bf bx}_{0, j_2, r, 1, R_i}]
\chi_{sub}[{{\vec b_1}}] \right) \cr
&+ \sum_{j_2, r', M, R_i} I^{L}_{j_2, \hat r, M, R_i} \chi_{sub}[{\vec c_0}] }}
Where $\vec b_{0,1}$ are  given by \defqbc\ with $R_1=0,1$ respectively
 and $\vec c_0$ is
given by \defqbc\ with $j_1=0$, $r = \hat r$, $R_1 = M$.
The quantities $n[{\bf xx}_{j_1, j_2, r, R_i}]$ in \fineq\ are the number
of copies of the irreducible representation, with listed quantum
numbers, that appear  in $A$, and $I^{L}_{j_2, \hat r , M, R_i}$ are the
indices \firstind\ made out of these numbers.

Notice that most of the discussion in this section goes through unchanged if
we were to consider the supergroup $SU(2|4)$  (or $SU(2|m)$). The representation theory
of this group  was studied in \refs{\mark,\kim}   and the
index was used in \hailin\ to analyze various field theories with this symmetry.
The index for the plane wave matrix model is given by an expression like (4.3) below
but
without the denominators (this is then inserted into (4.1)).
Notice that the fact that the index for $N=4$ Yang Mills and the index for
the plane wave matrix model are different implies that we cannot continuously
interpolate between ${\cal N}=4$ super Yang Mills and the plane wave matrix model while
preserving the $SU(2|4)$ symmetry.
In \lubos\ BPS representations and an index for $SU(1|4)$ were considered.

\subsec{The Witten Index ${\cal I}^{WR}$ }
As in Section 2, we may define a second index ${\cal I}^{WR}$. The theory
for this index is almost identical. We focus on the supercharge,
$\bar{Q}_{m-1}^{- \half }$ which has $SU(2) \times SU(2)$ quantum
numbers, $(j_1^3 , j_2^3) = (0,- \half )$, $r = -1$ and $SU(4)$ quantum numbers
$(0,0,\ldots,1)$. Let $\bar{S} = \bar{Q}^{\dagger}$.

It is then easy to verify (see Appendix A) that \eqn\qsbcom{ 2
\{\bar{S}, \bar{Q} \}= H-2J_2-2\sum_{k=1}^{m-1} {k \over
m}R_k+{(4-m)r \over 2m} = E - (E_2-2)  \equiv \bar{\Delta}.} It follows
from \qsbcom\ that every state in a unitary representation of the
superconformal group has $\bar{\Delta} \geq 0$.

Following \winddef\
we define \eqn\rwinddef{{\cal I}^{WR}=Tr_R \left[ (-1)^F \exp(-\beta
\bar{\Delta} + \bar{M}) \right],} where $\bar{M}$ is the part of the
superconformal algebra that commutes with $\bar{Q}$ and $\bar{S}$.

The Cartan elements of this subalgebra are given in terms of those of the
algebra by
\eqn\qnfirstright{E'=E+j_2, ~~~j'_1=j_1,~~~ r'
={(m-1) (r +R_{m-1}) \over m} + \sum_{p=1}^{m-2}{p \over m} R_p, ~~~R_k'=R_{k}.}
Note that $r'$ depends on $r$ and $R_{m-1}$ only through the combination
$r + R_{m-1}$.
We then find that the index \rwinddef\ is zero on long representations and
for ${\bf c}, {\bf b}$ representations it is equal to
\eqn\indexcharR{ \eqalign{
{\cal I}^{WR}[{\bf bx}_{j_1, 0,r,R_i}] =& \chi_{sub}[\vec {\bar b}]
\cr
{\cal I}^{WR}[{\bf cx}_{j_1, j_2, r,
R_i}]=& (-1)^{2 j_2+1} \chi_{sub}[{\vec {\bar c}}]
}}
where the representation of the subalgebra is specified by the vectors $\vec{{\bar b}}$, $\vec{\bar{c}}$ in the basis $[E',j_1',r',R_k']$ specified by \qnfirstright.
\eqn\defqbcR{ \eqalign{
 \vec{{\bar b}} & = [ - {3\over 2} r+2r',j_1,r'(r + R_{m-1},R_k),R_k], \cr
 \vec{\bar c} & =[ 3 +{3} (j_2 - r/2) + 2r',j_1,r'(r + R_{m-1},R_k),R_k].
}}
where $r'$ is the function in \qnfirstright .
We find that on a general
representation (not necessarily irreducible) of the superconformal
algebra, ${\cal I}^{WR}$ evaluates to \eqn\fineqR{ \eqalign{
{\cal I}^{WR}[R]=&\sum_{j_1, r, R_i} \left( n[{\bf xb}_{j_1, 0, r, R_k,0}]
\chi_{sub}[{{\vec{\bar{b}}_0}}] + n[{\bf xb}_{j_1, 0, r, R_k,1}]
\chi_{sub}[{{\vec{\bar{b}}_1}}] \right) \cr &+ \sum_{j_1, r'', R_k,N}
I^{R}_{j_1, r'',R_k,N} \chi_{sub}[{\vec{\bar{c}}_0}] }}
Where $\vec{\bar{b_{0,1}}}$ are  given by \defqbcR\ with $R_{m-1}=0,1$ respectively and $\vec{\bar{c_0}}$ is
given by \defqbcR\ with $j_2=0$, $r = r''$, $R_{m-1} = N$.
The quantities $n[{\bf xx}_{j_1, j_2, r, R_i}]$ in \fineq\ are the number
of copies of the irreducible representation, with listed quantum
numbers, that appear  in $R$, and $I^{R}_{j_1, r'',R_k,N}$ are the
indices \secondind\ made out of these numbers.

The main lesson we should extract from \fineq, \fineqR\ is that each
of the indices defined in section two are multiplied by {\it
different} $SU(1,2|m-1)$ (or $SU(2,1|m-1)$) characters in \fineq
\fineqR. This shows that the Witten indices \winddef\ \rwinddef\
capture
 all the protected that follows from the supersymmetry algebra alone.

\newsec{Computation of the Index in $\CN=4$ Yang Mills on $S^3$}
\subsec{Weak Coupling}
We will now evaluate the index \winddef\ for free $\CN=4$ Yang Mills
on $S^3$. In the Free theory this Index may be evaluated either by
simply counting all gauge invariant states with $\Delta=0$ and
specified values for other charges \refs{\sundborg, \sphere} or by
evaluating a path integral \sphere. The two methods give the same
answer. We will give a very brief description of the path integral
method, referring the reader to \sphere\ for all details. One
evaluates the path integral over the $\Delta=0$ modes of all the
fields of the $\CN=4$ theory on $S^3 \times S^1$ with periodic
boundary conditions for the fermions around $S^1$ (to deal with the
$(-1)^F$ insertion) and twisted boundary conditions on all charged
fields (to insert the appropriate chemical potentials). While the
path integral over all other modes may be evaluated in the one loop
approximation, the path integral over the zero mode of $A_0$ on this
manifold must be dealt with exactly (as the integrand lacks a
quadratic term for this mode, the integral over it is always
strongly coupled at every nonzero coupling no matter how weak). Upon
carefully setting up the problem one finds that the integral over
$A_0$ is really an integral over the holonomy matrix $U$, and the
index ${\cal I}^{WL}$ evaluates to \eqn\exacti{ {\cal I}_{YM} = \int [dU]
\, \exp{\left\{ \sum {1 \over m} f(t^m,y^m,u^m,w^m) {\rm
tr}(U^{\dagger})^m {\rm tr} U^m \right\} }} where $f(t, y, u,
w)$ is the index ${\cal I}^{WL}$ evaluated on space of `letters' or
`gluons' of the $\CN=4$ Yang Mills theory. As a consequence, in
order to complete our evaluation of the index \exacti\ we must
merely evaluate the single letter partition function $f$.

$f$ may be evaluated in many ways. Group theoretically, we note that
the letters of Yang Mills theory transform in the `fundamental'
representation of the superconformal group (the representation whose
quantum lowest weight state has quantum numbers $E=1$, $j_1=j_2=0$,
$R_1=R_3=0$ and $R_2=1$). $f$ is simply the supertrace over this
representation, which we have evaluated using group theoretic
techniques in Appendix C.

It is useful, however, to re-derive this result in a more physical
manner. The full set of single particle $\Delta=0$ operators in Yang
Mills theory is given by the fields listed in Table 2. below, acted
on by an arbitrary numbers of the two derivatives $\partial_{+ \pm}$
(see the last row of Table 2) modulo the single equation of motion
listed in the second last row of Table 2. \bigbreak
\bigskip
{
\offinterlineskip
\def\tablerule{\noalign{\hrule}}
\def\tableskip{\omit&height 3pt&&&&&\omit\cr}
\halign{\tabskip = .7em plus 1em
\vtop{\hsize=6pc\pretolerance = 10000\hbadness = 10000
\normalbaselines\noindent\it#\strut}
&\vrule #&#\hfil &\vrule #&# \hfil &\vrule#
&\vtop{\hsize=11pc \parindent=0pt \normalbaselineskip=2pt
\normalbaselines \rightskip=3pt plus2em #}\cr
\noalign{\hrule height2pt depth2pt \vskip3pt}
\multispan5\bf Table 2: Letters with ${\bf \Delta = 0}$ \hfil\strut\cr
\noalign{\vskip3pt} \tablerule
\omit&height 3pt&\omit&&\omit&&\omit \cr
\bf Letter && $\bf (-1)^F [E;j_1,j_2]$ && $ \bf [q_1,q_2,q_3] $&& $\bf [R_1,R_2,R_3]$ \cr
\tablerule
\tableskip $X,Y,Z$ && [1,0,0] && [1,0,0]+{\rm cyclic} && [0,1,0]+ [1,-1,1]+[1,0,-1]\cr
\tablerule
\tableskip $\psi_{+,0 ;-++}+{\rm cyc}$ && $-[{3 \over 2},{1 \over 2},0]$ &&$[-{\half},{\half},{\half}] + {\rm cyc}$&&$[1,-1,0],[0,1,-1],[0,0,1]$ \cr
\tableskip $\psi_{0,\pm,+++}$&&$-[{3 \over 2},0,\pm {1 \over 2}]$ &&$[{\half},{\half},{\half}]$&&$[1,0,0]$ \cr
\tablerule
\tableskip $F_{+ +}$ && $[2,1,0]$ &&$[0,0,0]$ && $[0,0,0]$ \cr
\tablerule
\tableskip $\partial_{++} \psi_{0,-;+++} + \partial_{+-}\psi_{0,+;+++} =0$&& $[{5 \over 2},{\half},0]$ && $[{\half},{\half},{\half}]$ && $[1,0,0]$ \cr
\tablerule
\tableskip $\partial_{+ \pm}$ && $[1,\half,{\pm \half}]$&&$[0,0,0]$ && $[0,0,0]$ \cr
\tableskip \tablerule \noalign{\vskip 2pt} \tablerule
}}
\bigskip

In Table 2 we have listed both the $SU(4)$ Cartan charges $R_1, R_2,
R_3$ used earlier in this paper, as well as the $SO(6)$ Cartan
charges, $q_1, q_2, q_3$ (the eigenvalues in each of the 3 planes of
the embedding $R^6$) of all fields.

To find $f$ we evaluate $\winddef$ by summing over the letters

\eqn\fdef{\eqalign{ f &= \sum_{\rm letters} (-1)^F t^{2(E+j_1)} y^{2
j_2} v^{R_2} w^{R_3} \cr &={t^2(v + {1 \over w} + {w \over v}) -
t^3(y + {1 \over y}) - t^4(w + {1 \over v} + {v \over w}) + 2 t^{6}
\over (1 - t^3 y)(1 - {t^3 \over y})}.}}  Remarkably the expression
for $1-f$ factorizes \eqn\oneminuf{ 1 - f = {(1 - t^2/w)(1 - t^2
w/v) (1 - t^2 v) \over (1 - t^3 y)(1 - t^3/y)}} The expression for
${\cal I}^{WL}_{YM}$ is well defined (convergent) only if $t, y, v, w$ have
values such that every contributing letter has a weight of modulus
$< 1$; applying this criterion to the three scalars and the two
retained derivatives yields the restriction $t^2 v <1,~~~t^2 /w <1,
~~~t^2 v/w <1 ,~~~ t^3y<1, t^3/y>1$. It follows from \oneminuf\ that
$f<1$ for all legal values of chemical potentials.

We will now proceed to evaluate the integral in \exacti, using
saddle point techniques, in the large $N$ limit (note, however, that
\exacti\ is the  {\it exact} formula valid for all $N$). To process this
formula, we convert the integral over $U$ to an integral over its
$N^2$ eigenvalues $e^{i \theta_j}$. We can conveniently summarize
this information in a density distribution $\rho(\theta)$ with:
\eqn\rhonorm{ \int_{-\pi}^{\pi} d \theta \, \rho(\theta) = 1 }

This generates an effective action for the eigenvalues given by
\sphere\ \eqn\distact{\eqalign{ S[\rho(\theta)] &= N^2  \int d
\theta_1 \int d \theta_2 \rho(\theta_1) \rho(\theta_2) V(\theta_1 -
\theta_2) = \cr
  &= {N^2 \over 2 \pi} \sum_{n=1}^\infty |\rho_n|^2 V_n (T),
}} with \eqn\potmodes{\eqalign{ V_n &= {2 \pi \over n} (1-
f(t^n,y^n,u^n,w^n)), \cr \rho_n &= \int d \theta \rho(\theta) e^{i n
\theta } .}} As $(1 - f)$ is always positive for all allowed values
of the chemical potential, it is clear that the action \distact\ is
minimized by $\rho_n = 0, n > 0$; $\rho_0 = 1$. The classical value
of the action vanishes on this saddle point, and the Index is given
by the gaussian integral of the fluctuations of $\rho_n$ around zero.
This allows us to write \eqn\ymindex{ \left. {{\cal I}^{WL}_{YM}}\right|_{N=\infty}
= \prod_{n =
1}^{\infty} {1 \over 1 - f(t^n,y^n,v^n,w^n)} .}
If we think about the 't Hooft limit of the theory it is also interesting
to compute the index over single trace operators.
This is given by
\eqn\singtra{\eqalign{
Z_{s.t.} = & - \sum_{r=1}^\infty  { \varphi(r) \over r}
\log\left[1 - f( t^r, y^r, v^r, w^r) \right]
\cr
= &   {t^2/w
\over 1 - t^2/w} + {v t^2 \over 1 - v t^2} + {t^2 w/v \over 1 - t^2
w/v} - {t^3/y \over 1 - t^3/y} - {t^3 y \over 1 - t^3 y}
}}
where $\varphi$ is the Euler Phi function and we used that
$\sum_{r} {\varphi(r) \over r} \log(1-x^r) = {-x \over 1-x} $.
The result \ymindex\ is simply the multiparticle contribution that we get from
\singtra .

Note that the action \distact\ vanished on its saddle point; as a
consequence \ymindex\ is independent of $N$ in the large $N$ limit.
This behavior, which is is reminiscent of the partition function of
a large $N$ gauge theory in its confined phase, is true of \ymindex\
at all finite values of the chemical potential. In this respect the
index ${\cal I}^{YM}$ behaves in a qualitatively different manner from the
free Yang Mills partition function over supersymmetric states (see
the next section).  This partition function displays confined
behavior at large chemical potentials (analogous to low
temperatures) but deconfined behavior (i.e. is of order $e^{N^2}$)
at small chemical potentials (analogous to high temperature). It
undergoes a sharp phase transition between these two behaviors at
chemical potentials of order unity. Several recent studies of Yang
Mills theory on compact manifolds have studied such phase
transitions, and related them to the nucleation of black holes in
bulk duals \refs{\sundborg, \sphere, \hong, \spenta, \gl, \tloop,
\pallav, \torus}. The index ${\cal I}^{WL}_{YM}$ does not undergo this
phase transition, and is always in the `confined' phase. We
interpret this to mean that it never `sees' the dual supersymmetric
black hole phase.

At first sight we might think that this is a contradiction, since
the black holes give a large entropy. On the other hand we are
unaware of a clear argument which says that black holes should
contribute to the index. For example, it is unclear whether the
Euclidean black hole geometry should contribute to the path integral
that computes the index. While the Lorentzian geometry of the black
hole is completely smooth, if we compactify the Euclidean time
direction with periodic boundary conditions for the spinors, then
the corresponding circle shrinks to zero size at the horizon, which
would represent a kind of singularity. See the next section and
Appendix D for a mechanism for how this phenomenon (the excision of
the black hole saddle point) might work in Lorentzian space.

We now present the expression for the Index in a new set of
variables that are more symmetric, and for some purposes more
convenient, in the study of Yang Mills theory. We will use these
variables in the next section. Let us choose to parameterize charges
in the subalgebra by \eqn\newvar{J_2,~L_1 = E + q_1 - q_2 - q_3,~L_2
= E + q_2 - q_1 - q_3,~L_3 = E + q_3 - q_1 - q_2. } Note that $L_i$
are positive for all Yang Mills letters. A simple change of basis,
(see Appendix C) yields \eqn\omfrewr{ 1-f={(1-e^{-2 \g_1}) (1-e^{-2
\g_2}) ( 1- e^{-2 \g_3}) \over \left( 1-e^{- \zeta -\g_1-\g_2 -\g_3}
\right) \left( 1-e^{+\zeta -\g_1-\g_2 -\g_3} \right)}} where
\eqn\fefn{f =\sum_{\rm letters} (-1)^Fe^{\gamma_1 L_1 +\gamma_2 L_2
+\gamma_3 L_3 + 2\zeta j_2 }.}


In section six we will write an explicit exact formulas for the index \exacti\
 for  $\gamma_3 =\infty$.

Further studies on the spectrum of free Yang Mills can be found in
\refs{\morales,\bianchione,\bianchitwo,\bianchithree}.

\subsec{Strong Coupling} \subseclab\strcoup
According to the AdS/CFT correspondence, $\CN=4$ Yang Mills theory
on $S^3$ at large $N$ and large $\lambda$ has a dual description as
a weakly coupled IIB theory on the large radius $AdS_5 \times S^5$.
At fixed energies in the large $N$ limit, the spectrum of the bulk
dual is a gas of free gravitons, plus superpartners,  on
$AdS_5\times S^5$. In this subsection section we will compute the
index ${\cal I}^{WL}_{YM}$ over this gas of masssless particles, and find
perfect agreement with \ymindex.

Note that states with energies of order one do not always dominate
the partition function at chemical potentials of unit order. At
small values of the chemical potential, the usual partition function
of strongly coupled Yang Mills theory is dominated by black holes.
However, as we have explained in the previous subsection, we do not
see an argument for the black hole saddle point to contribute to the
Index, and apparently it does not.

We now turn to the computation. When the spectrum of (single
particle) supergravitons of Type $II B$ supergravity compactified on
$AdS_5 \times S^5$ is organized into representations of the
superconformal group, we obtain representations that are built on a
lowest weight state that is a $SU(2)\times SU(2)$ in the
$(n,0,0)_{SO(6)} = (0,n,0)_{SU(4)}$ representation of the R-symmetry
group \gunmar .
 The representation with $n=1$ is the Yang-Mills multiplet.
The representation with $n=2$ is called the 'supergraviton'
representation. These representations preserve 8 of 16
supersymmetries. In the language of section 2, they are of the form
${\bf b b}$. When restricted to $\Delta = 0$, they yield a
representation of the subalgebra that we shall call $S_n$. $S_n$ has
lowest weights $E'=n, j_2=0, R_2=n, R_3=0$. The states of $S_n$ are
tabulated explicitly in appendix C. The state content of $n=1$ is
somewhat different and is tabulated separately. This can also be
found by looking at the list of Kaluza Klein modes in \gunmar .

The index on single-particle states may now be calculated in a
straightforward manner. The supercharacter of $S_n$ may be read off
from the appendix and is given by \eqn\snsupchar{\chi_{Sn} = {(t^{2
n} \chi^{SU(3)}_{n,0}(v,w) - t^{2n + 1} \chi^{SU(3)}_{n-1,0}(v,w) (y
+ 1/y) + \ldots) \over  (1 - t^3 y)(1 - t^3/y)}.} The $SU(3)$
character that occurs above is described by the Weyl Character
Formula described in the Appendix C. To obtain the index, we simply
need to calculate \eqn\Isp{{\cal I}_{sp} = \sum_{n=2}^{\infty}
\chi_{Sn} + \chi_{S1}.} The sums in \Isp\ are all geometric and are
easily performed,  yielding the single particle contribution
\eqn\zstbps{ {\cal I}_{sp} = {t^2/w
\over 1 - t^2/w} + {v t^2 \over 1 - v t^2} + {t^2 w/v \over 1 - t^2
w/v} - {t^3/y \over 1 - t^3/y} - {t^3 y \over 1 - t^3 y} }
This matches precisely \singtra .

To obtain the index for the Fock-space of gravitons we use the formula,
 justified in the Appendix C,  that relates the index of one particle to the index
of the Fock Space. \eqn\sugraindex{ \eqalign{ {{\cal I}^{WL}_{grav}} &=
\exp \left[\sum_n { 1 \over n}  {\cal I}_{sp} [t^n,v^n,w^n,y^n] \right] \cr
  &= \prod_{n=1}^\infty  {(1 - t^{3 n}/y^n)
   ( 1 - t^{3 n} y^n) \over (1 - t^{2 n}/w^n) (1 - v^n t^{2n}) (1 - t^{2 n} w^n/v^n)}
}} in perfect agreement with \ymindex.

Finally, let us point out that the value of the index is the same in ${\cal N}=1$ marginal
deformations of ${\cal N}=4$.\foot{
These theories have  the superpotential $Tr[ e^\beta \phi_1 \phi_2 \phi_3 -
 e^{-\beta} \phi_1 \phi_3
\phi_2 + c (\phi_1^3 + \phi_2^3 +\phi_3^3 ) ]$. If $c$ is nonzero, then we should set all
chemical potentials $\gamma_i$ to be equal in the original ${\cal N}=4$ result, since
we loose two of the $U(1)$ symmetries. }

\newsec{The partition function over BPS states}
In this section we will compute the partition function over BPS
states that are annihilated by $Q$ and $S$ in $\CN=4$ Yang Mills at
zero coupling and strong coupling. We perform the first computation
using the free Yang Mills action, and the second computation using
gravity and the AdS/CFT correspondence, together with a certain
plausible assumption. Specifically,  we assume that the
supersymmetric density of states at large charges is  dominated by
the supersymmetric black holes of \refs{\reallone, \reallthree,
\cvetic}.

At small values of chemical potentials (when these supersymmetric
partition functions are dominated by charges that are large in units
of  $N^2$) we find that these partition functions are qualitatively
similar at weak and strong coupling but differ in detail, in these
two limits. Moreover, each of these partition functions differs
qualitatively from index computed in the previous section.

Before turning to the computation, it may be useful to give a more
formal description of the BPS states annihilated by $Q$ and $S$.
$Q$ may formally be thought of as an exterior derivative $d$, its
Hermitian conjugate $S$ is then $d^*$ and $\Delta$ is the Laplacian
$d d^* + d^* d$. States with $\Delta=0$ are harmonic forms that,
according to standard arguments (see \ref\gsw{
  M.~B.~Green, J.~H.~Schwarz and E.~Witten,
  ``Superstring Theory. Vol. 2: Loop Amplitudes, Anomalies And Phenomenology,'' Chapter 12, Cambridge University Press (1987). }, those arguments may all be reworded in the language
of $Q$ and $S$ and Hilbert spaces) are in one to one correspondence
with the cohomology of $Q$. ${\cal I}^{WL}$, the $(-1)^{degree}$ weighted
partition function over this cohomology is simply the (weighted)
Euler Character over this cohomology.

\subsec{Partition Function at $\Delta=0$ in Free Yang Mills}
Let \eqn\dzpf{Z_{free} =Tr_{\Delta=0} \left[ x^{2 H} e^{\mu_1 q_1 + \mu_2 q_2 +
\mu_3 q_3 + 2 \zeta J_2} \right] } where $x= e^{-\beta \over 2}$, and $q_1,
q_2, q_3$ correspond to the $SO(6)$ Cartan charges (related to $R_1, R_2, R_3$
by the formulas in Appendix C).  In Free Yang Mills theory this
partition function is easily computed along the lines described in
subsection 4.1; the final answer is given by the formula
\refs{\sundborg, \sphere}
 \eqn\pff{Z=\int DU \exp \left[ \sum_n
\left( f_B(x^n,n \mu_i, n \zeta) + (-1)^{n+1} f_F(x^n, n \mu_i, n
\zeta) \right)  {Tr U^n Tr U^{-n} \over n } \right] } where $U$ is a
unitary matrix and the relevant `letter partition functions' are
given by
\eqn\fbosonic{ \eqalign{ f_B&= {(e^{\mu_1} + e^{\mu_2} +
e^{\mu_3})x^2+ x^4 \over (1-x^2 e^{{\zeta }}) (1-x^2 e^{-{\zeta }})}  \cr
f_F&={ x^3} ( 2 \cosh {\zeta }  e^{ {\mu_1+\mu_2 +\mu_3 \over 2}} +
e^{{\mu_1 + \mu_2 -\mu_3 \over 2}}+  e^{{\mu_1 -\mu_2 +\mu_3 \over 2}}
+e^{{-\mu_1 + \mu_2 +\mu_3 \over 2}} )  -x^{5}e^{ {\mu_1+\mu_2 +\mu_3 \over 2}} \over (1-x^2 e^{{\zeta}}) (1-x^2 e^{-{\zeta}})}.}

As explained in the previous section, \pff\ and \fbosonic\ describe
a partition function that undergoes a phase transition at finite
values of chemical potentials. For chemical potentials such that
$f_B+f_F<1$, the integral in \pff\ is dominated by a saddle point on
which $|TrU^n|=0$ for all $n$. In this phase the partition function
is obtained from the one loop integral about the saddle point (as in
section 4.1) and is independent of $N$ in the large $N$ limit. The
density of states in this phase grows exponentially with energy,
$\rho(E) \propto e^{\beta_H E}$ where $\beta_H=-\ln({7-3\sqrt{5}
\over 2})=1.925$ and the system undergoes a phase transition when
the effective inverse temperature becomes smaller than $\beta_H$
(e.g., on setting all other chemical potentials to zero, this
happens at $x=e^{-\beta_H \over 2}$).

At smaller values of chemical potentials  \pff\ is dominated by a
new saddle point. In particular, in the limit $\zeta \ll 1$ and
$\beta \ll 1$, the integral over $U$ in \dzpf\ is dominated by a
saddle point on which $Tr U^n Tr U^{-n}=N^2$ for all $n$, the
partition function reduces to \eqn\pfz{\ln Z= N^2 \sum_n {1 \over n}
\left[ f_B(x^n,n \mu_i, n \zeta) + (-1)^{n+1} f_F(x^n, n \mu_i, n
\zeta) \right].}

In the rest of this subsection we will, for simplicity, set
$\mu_1=\mu_2=\mu_3= \mu$ and thereby focus on that part of
cohomology with $q_1=q_2=q_3 \equiv q $. The relevant letter
partition functions reduce to
\eqn\fbosonica{ \eqalign{ f_B&= {3
e^{\mu}x^2+ x^4 \over (1-x^2 e^{{\zeta}}) (1-x^2 e^{-{\zeta }})  }  \cr
f_F&={ \left(e^{ {3 \mu \over 2}}(2 \cosh \zeta -x^2) + 3 e^{{\mu
\over 2}}\right) x^{3} \over (1-x^2 e^{{\zeta }}) (1-x^2
e^{-{\zeta}})} }} In the limit $\beta \ll 1$, $\zeta \ll 1$ \pfz\
reduces to \eqn\pfrel{\ln Z= N^2 {1 \over (\beta^2 - \zeta^2)}
{f(\mu)} } where \eqn\fdef{ f(\mu) = \left( \zeta(3) + 3 Pl(3,
e^\mu) - 3 Pl(3,-e^{{\mu \over 2}})  -Pl(3, -e^{{3 \mu \over
2}})\right)} and the PolyLog function is defined by \eqn\polylogdef{
Pl(m, x) = \sum_{n=1}^\infty {x^n \over n^m} } This partition
function describes a system with energy $E$, angular momentum $j_2$,
$SO(6)$ charge $(q, q, q)$ and entropy $S$ given by\foot{Physically,
the equations below describe Free Yang Mills theory at fixed values
of charges in the limit $T \to 0$ ($T$ is the temperature). In the
free theory this limit retains only supersymmetric states at all
values of charges. On the other hand the black holes in
\refs{\reallone, \reallthree, \cvetic} are supersymmetric in the same
limit only for a subfamily of charges. See the next section for more
discussion on this puzzling difference. }

\eqn\en{\eqalign{ {2 j_1 \over N^2 } \sim {E \over N^2}&=2 { \beta f(\mu) \over (\beta^2
-\zeta^2)^2}, \cr {2 j_2 \over N^2} &= 2 {   \zeta f(\mu) \over (\beta^2
-\zeta^2)^2} \cr {q \over N^2}&= {g(\mu)\over \beta^2-\zeta^2} \cr
{S \over N^2}&={ 3 f(\mu) -\mu g(\mu) \over \beta^2 -\zeta^2} }}
where \eqn\gdefini{g(\mu)={f'(\mu) \over 3} =  \left( Pl(2, e^\mu )-
{1\over 2} Pl(2, -e^{{\mu \over 2}}) - {1 \over 2} Pl(2, -e^{{3 \mu
\over 2}}) \right) . }
We see that for high temperatures, this partition function looks like
a gas of massless particles in 2+1 dimensions. Note that in this limit
$E \sim 2 j_1 \gg q $.

We will sometimes be interested in the partition function with only
those chemical potentials turned on that couple to charges that
commute with $Q$ and $S$. This is achieved if we choose $\mu
={ \beta \over 3}$. In the limit $\beta \ll 1$, $\zeta \ll 1$
we have $\mu \ll 1$ and the partition function and charges are given
by \pfrel\ and \en\ with $\mu \sim 0$; note that $f(0)=7
\zeta(3) $ and $g(0) ={ \pi^2 \over 4}$.

Note that, although the index ${\cal I}^{WL}_{YM}$ and the cohomological
partition function $Z_{\rm free}$ both traces over $Q$ cohomology ,
the final results for these two quantities in Free Yang Mills theory
are rather different. For instance, at finite but small values of
chemical potentials, $\ln Z_{\rm free}$ is proportional to $N^2$
(see \pfrel) while ${\cal I}^{WL}_{YM}$ is independent of $N$ (see
\ymindex). Clearly cancellations stemming from the fluctuating sign
in the definition of ${\cal I}^{WL}_{YM}$ cause the index to see a smaller
effective number of states. In Appendix D we explain, in more
detail, how this might come about.

\subsec{Cohomology at Strong Coupling: Low energies }

We now turn to the study of $Q$ cohomology at strong coupling and
low energies. In this limit the cohomology is simply that of the
free gas of supergravitons in $AdS_5 \times S^5$, and may be
evaluated following the method of subsection subsection \strcoup. We
will calculate the quantity \eqn\trace{ Z = {\rm Tr}\left[x^{2 H}
z^{2 J_1} y^{2 J_2} v^{R_2} w^{R_3} \right] } over the supergraviton
representations restricted to states of $\Delta = 0$. We recall that
the single particle states form an infinite series of short reps of
the $N=4$ superconformal algebra where the primary is a lorentz
scalar with energy $n$ with R-charges $[0,n,0]$.

The trace over single particle states may be easily calculated. The
answer is \foot{In the notation of the
previous subsection, with $y=e^\zeta$, \eqn\singleparthree{ \eqalign{ Z^{\rm res}_{sp} &=
Tr\left[x^{2 H} y^{2 J_2} u^{2 \sum q_i}\right]= {{\rm num^{\rm
res}_{bos}+num^{\rm res}_{fer}} \over {\rm den^{\rm res}} } \cr {\rm
den} &= (1 - x^2 u^2)^3 (1 - x^2/y) (1 - x^2 y) \cr {\rm num_{fer}}
&= 3 u x^3 - 2 u^3 x^5 + 3 u^5 x^7 + (u^3 x^3)/y + (u^3 x^7)/y \cr
&+ u^3 x^3 y + u^3 x^7 y \cr {\rm num_{bos}} &= 3 u^2 x^2 + x^4 - 3
u^4 x^4 + 2 u^6 x^6 + u^6 x^{10} \cr &+ (3 u^4 x^6)/y - (u^6 x^8)/y
+ 3 u^4 x^6 y - u^6 x^8 y }} }
 \eqn\singlepart{ \eqalign{ Z_{\rm sp} &= {{\rm
num_{bos}+num_{fer}} \over {\rm den} } \cr {\rm den} &= (1 - x^2/w)
(1 - x^2 v) (1 - x^2 w/v) (1 - x^2 z/y)(1 - x^2 z y) \cr {\rm
num_{fer}} &= x^3/y + x^3 y + x^3 z/v + v x^3 z/w +  w x^3 z - 2
x^5 z \cr &+ v x^7 z +  x^7 z/w + w x^7 z/v +  x^7 z^2/y +  x^7
 z^2 y \cr {\rm num_{bos}} &= v x^2 + x^2/w + w x^2/v - x^4/v - v
x^4/w -   w x^4\cr&+ 2  x^6 + x^6 z/(y v) + v x^6 z/(w y) +  w
x^6 z/y -  x^8 z/y \cr &+ x^6 z y/v + v x^6 z y/w +  w x^6 z y
-  x^8 z y +  x^4 z^2 +  x^{10} z^2 }}
 The full (multi particle) partition
function over supersymmetric states may be obtained by applying  the
formulas of Bose and Fermi statistics to \singlepart.

Special limits of \singlepart\ will be of interest in the next
section. For instance, the limit $z \rightarrow 0$ focus on states
with $\Delta=0$ and $j_1=0$, i.e.  $(1/8)$ BPS states. In this limit
\singlepart\ becomes \eqn\oneeigth{ \eqalign{ Z^{1/8}_{\rm bos-sp} &=
{1 - (1 - x^2/w)(1 - v x^2 )(1 - w x^2/v) + x^6 \over (1 - x^2/w)(1
- v x^2)(1 - w x^2)} \cr
 Z^{1/8}_{\rm fer-sp} &= {x^3 (y + 1/y) \over (1 - x^2/w)(1 - v x^2) ( 1 - w x^2/v)}
}} In terms of the $\gamma_i$ variables introduced at the end of
subsection 4.1  \eqn\oneeigthgamma{ \eqalign{ Z^{1/8}_{\rm bos-sp} &= {1 - (1
- e^{-2 \gamma_1})(1 - e^{-2 \gamma_2})(1 - e^{-2 \gamma_3}) + e^{-2
(\gamma_1 + \gamma_2 + \gamma_3)} \over (1 - e^{-2 \gamma_1})(1 -
e^{-2 \gamma_2})(1 - e^{-2 \gamma_3})} \cr Z^{1/8}_{\rm fer-sp} &=
{e^{-\gamma_1 - \gamma_2 - \gamma_3} \left[ e^{\zeta} + e^{-\zeta}
\right] \over (1 - e^{-2 \gamma_1})(1 - e^{-2 \gamma_2})(1 - e^{-2
\gamma_3})} }}

Applying the formulas for Bose and Fermi statistics, it is now easy to see that the partition function over the Fock space, including multi-particle states, is given by
\eqn\parteigthstrong{
Z^{1/8}(\zeta, \gamma_1, \gamma_2,\gamma_3)=\prod_{n, m, r=0}^\infty
 { \prod_{s=\pm 1} (1 + e^{ s \zeta} e^{-(2 n +1) \gamma_1-(2 m +1) \gamma_2 - (2 r +1) \gamma_3})\over (1 - e^{ -  2 n \gamma_1- 2 m \gamma_2 - 2 r \gamma_3 } ) ( 1 - e^{-(2 n +2) \gamma_1- (2m+2)\gamma_2 - (2 m +2)\gamma_3})
 }}

Finally, in order to compute the rate of growth of the cohomological
density of states with respect to energy, we set  $z,y,v,w
\rightarrow 1$. This gives the ``blind'' single particle partition
function which is \eqn\zblind{\eqalign{ Z^{\rm bl}_{\rm bos-sp} &= {x^2
(3 - 2 x^2 + 8 x^4 - 2 x^6 + x^8) \over (1 - x^2)^5} \cr Z^{\rm
bl}_{\rm fer-sp} &= {x^3 (5 - 2 x^2 + 5 x^4) \over (1 - x^2)^5} }}

The full partition function is given by \eqn\zfullblind{ Z^{\rm bl}
= \exp\left[\sum_n {Z^{\rm bl}_{\rm bos-sp}(x^n) + (-1)^{n+1} Z^{\rm
bl}_{\rm fer-sp}(x^n)  \over n} \right]} Let \eqn\thetadef{x =
e^{-\beta \over 2}.} At small $\beta$ this partition function is
approximately given by \eqn\pfa{\ln Z= {63 \zeta(6) \over 4
\beta^5}.} It follows that the entropy as a function of energy is
given by
\eqn\enteff{S(E)=h\log n(E) \sim {6\over 5} \left({315 \zeta(6)
\over 4} \right)^{{1\over 6}} E^{5/6} \approx 2.49 E^{5/6}.} Note that this is slower
than the exponential growth of the same quantity at zero coupling.

\subsec{Cohomology at Strong Coupling: High Energies }
Gutowski and Reall \refs{\reallone,\reallthree}, and Chong, Cvetic,
Lu and Pope \cvetic\ have found a set of supersymmetric black holes
in global $AdS_5 \times S^5$, that are annihilated by the
supercharges $Q$ and $S$. These black holes presumably dominate the
supersymmetric cohomology at energies of order $N^2$ or larger. In
this subsection we will translate the thermodynamics of these
supersymmetric black holes to gauge theory language, and compare the
results with the free cohomology of subsection 5.1.

Restricting to black holes with $q_1=q_2=q_3=q$ these solutions
constitute a two parameter set of solutions, with thermodynamic
charges, translated to Yang Mills Language via the AdS/CFT
dictionary\foot{ We have set $g =1$ in \cvetic\ and set $E_{CFT} =
E_{Chong ~et ~al}/G_5$, where $G_5 = G_{N 5}/R^3_{AdS}$ is the value
of Newton's constant in units where the $AdS_5$ radius is set to
one. $S_{here} = S_{Chong~et~al}/G_5$. For ${\cal N}=4$ Yang Mills
we have $G_5 = { \pi \over 2 N^2} $. To convert formulas in
\refs{\reallone,\reallthree} simply set this value for the five
dimensional Newton constant in their expressions.},
 \eqn\cce{ \eqalign{ {E
\over N^2}&={(a + b)} {\left[ (1-a)(1-b) +(1 +a)(1+b) (2-a-b)\right]
\over {2 (1 -a )^2 (1-b)^2} } \cr  {j_1+j_2 \over N^2}&= { { (a +b)
(2a +b + a b)} \over {2 (1-a )^2 (1-b)} } \cr {j_1-j_2\over N^2}&={
{ (a +b) (a + 2b + ab)} \over {2 (1-a )(1-b)^2} } \cr {q \over
N^2}&={ (a + b)  \over {2  (1 -a )(1 - b )}} \cr {S \over N^2}&= { {
\pi (a +b) \sqrt{a + b + a b }} \over { (1-a) (1- b)} }. \cr  }}
Setting $a=1-(\beta'+\zeta')$ and $b=1-(\beta'-\zeta')$, and
assuming $\beta' \ll 1$, $\zeta' \ll 1$, \cce\ reduces to \eqn\cces{
\eqalign{ { 2 j_1 \over N^2 }
\sim  {E \over N^2}&\sim  {8 \beta' \over (\beta^{'2} -\zeta^{'2})^2
} \cr {2 j_2 \over N^2}&\sim {- 8 \zeta' \over (\beta^{'2} -\zeta^{'2}
)^2 } \cr {q \over N^2}&\sim { 1 \over {\beta^{'2}-\zeta^{'2}} }  \cr {S
\over N^2}&\sim {2 \sqrt{3}  \pi \over \beta^{'2}-\zeta^{'2} }
 }}

Equations \cce\ and \en\ are have some clear similarities\foot{This
observation has also been made by H. Reall and R. Roiban.} in form,
but also have one important qualitative difference. \en\ has one
additional parameter absent in \cce. After setting the three $SO(6)$
charges equal the $Q$ cohomology is parametrized by 3 charges,
whereas only a two parameter set of supersymmetric black hole
solutions are available.

We should emphasize that in the generic, non BPS,  situation black hole
solutions are available for all values of the 4 parameters $q, j_2,
j_1$ and $E$ \cvetic . It is thus possible  to continuously lower the black
hole energy while keeping $q, j_2$ and $j_1$ fixed at arbitrary
values. The temperature of the black hole decreases as we lower its
energy, until it eventually goes to zero at a minimum energy.
However the extremal black hole thus obtained is supersymmetric (its
mass saturates the supersymmetric bound)  only on a 2 dimensional
surface in the 3 dimensional space of charges parameterized by $q,
j_2$ and $j_1$. For every other combination of charges the zero
temperature black holes are not supersymmetric (their mass is larger
than the BPS bound). We are unsure how this should be
interpreted\foot{Note that our Index ${\cal I}^{WL}_{YM}$, when specialized to states
with $q_1=q_2=q_3$, also depends on two rather than 3 parameters.}.
It is possible that, for other combinations of charges, the
cohomology is captured by as yet undiscovered supersymmetric black
solutions.

In order to compare the cohomologies in \en\ and \cce\ in more
detail, we choose $\mu$ in \en\ so that the equations for $E/N^2$
and $q/N^2$ in \en\ and \cce\ become identical (after a rescaling of
$\beta'$ and $\zeta'$). This is achieved provided that \eqn\chempeq{
f(\mu_c)^2= 16 g(\mu_c)^3} This equation is easy to solve
numerically. We find $\mu_c= -0.50366 \pm .00001$ and that
$f(\mu_c)=5.7765$, $g(\mu_c)=1.2776$. Plugging in $\mu=\mu_c$ into
the entropy formula in \en\ we then find \eqn\rat{ { S_{\rm Field}
\over S_{\rm Black-Hole}}= {3 {f(\mu) \over g(\mu)} -\mu \over 2
\sqrt{3} \pi } = 1.2927}

Another way to compare \en\ and \cce\ is the following. First notice that the
charge $q$ is much smaller than the energy in this limit, $q \ll E$.
Let us  set $\mu = \beta/3$ which is the value that we have in the index (though
we do not insert the $(-1)^F$ we have in the index).
 Since we are taking the
limit where $\beta$ is small we can evaluate $f$ in \en\ at zero, $f(0) = 7 \zeta(3)$.
By comparing the energies and entropies in \en\ and \cce\ and writing the free
energy as $E = N^2 c \beta^{-3}$, where $c$ is a ``central charge'' that measures the
number of degrees of freedom. Then we can compute
\eqn\ratios{
{ c_{\rm gravity} \over c_{\rm free-field-theory} } = { \pi^3 \over  14 \zeta(3) 3^{3/2} } \sim
0.35458...
}
It is comforting that this value is lower than one since we expect that interactions
would remove BPS states rather than adding new ones.
A similar qualitative agreement between the weak and strong coupling was
observed between
 the high temperature limit of uncharged black holes and the free Yang Mills theory
 \igor , where the ratio \ratios\ is 3/4. Note that for $\mu = \beta/3$  we can
 approximate $g$ in the expression for the charge in \en\ by $g(0) \not =0$. This
 agrees qualitatively with the expression coming from black holes.

\subsec{Cohomology at Intermediate Energies: Giant Gravitons and Small
Black Holes} Let us set $j_2=0$ or $a=b$ in \cce. We then expand the
resulting expression for low values of $a$. \eqn\ccelow{ \eqalign{
{E \over N^2}&\sim  3 a \sim 3 { q \over N^2}
 \cr
 {j_1  \over N^2}&\sim  3 a^2
\cr {S \over N^2}&\sim  {
2 \pi \sqrt{2} a^{3/2}  } . \cr
}}

It is possible to count the entropy of these black holes using D-branes in
$AdS$. This is not the same problem as counting them in the field theory, but perhaps
these results might be a good hint for the kind of states that we should look at
in the field theory.

In the small charge limit the black holes looks very similar to the black holes
that appear in toroidal compactifications of type IIB on $T^5$. Let us recall
first how the entropy of these black holes is counted \ascv .  We view
the black holes  as arising from two sets of intersecting
D3 branes $n_3 $ and $n'_3$ which intersect along a circle which is one of the
circles of $T^5$. One can then add momentum $L_0$
 along this circle. Their entropy is given by $S = 2\pi \sqrt{n_3 n'_3 L_0}$.
 This entropy arises as follows. Let us focus on the $T^4$ that is orthogonal
 to the common circle. The D3 branes can form any homomorphic surface on this $T^4$.
 The number of complex moduli of these surfaces goes as $n_3 n'_3$. There is an
 equal number of Wilson lines and there are $4 n_3 n'_3$ fermions. This gives
 central charge $c=6 n_3 n'_3$ and then using the Cardy formula we get the entropy.

We will now repeat the same counting for small black holes in our context.
First we recall that the theory contains giant graviton D3 branes which
can carry some of the charge.
Let us recall the description in \ref\MikhailovYA{
  A.~Mikhailov,
  ``Giant gravitons from holomorphic surfaces,''
  JHEP {\bf 0011}, 027 (2000)
  [arXiv:hep-th/0010206].
} for giant gravitons on the 5-sphere. We take
an arbitrary holomorphic 2-complex dimensional surface in $C^3$ and we intersect
it with $\sum |z_i|=1$. This gives a 3-real dimensional surface on $S^5$ which
will be a giant graviton. Let us focus first on surfaces that are invariant
under $\psi$ translations, where $\psi$ is an angle that rotates all $
z_i \to e^{i \psi } z_i$. The holomorphic surface in $C^3$ is specified by a
homogeneous polynomial of degree $n$.
\eqn\giantfo{
\sum_{n_1 + n_2 + n_3 = n}  C_{n_1,n_2, n_3} z_1^{n_1} z_2^{n_2} z_3^{n_3} =0
}
Think of $S^5 $ as an $S^1$ fibration on $CP^2$. Then \giantfo\ defines a holomorphic
surface in $CP^2$ and the resulting giant graviton on $S^5$ consists of this surface plus
the $S^1$ fiber which is parametrized by the angle $\psi$. For example, the maximum
size giant graviton that wraps an $S^3$ in $S^5$ \ref\McGreevyCW{
  J.~McGreevy, L.~Susskind and N.~Toumbas,
  ``Invasion of the giant gravitons from anti-de Sitter space,''
  JHEP {\bf 0006}, 008 (2000)
  [arXiv:hep-th/0003075].
} corresponds to the equation $z_1=0$.
The number of complex parameters characterizing the curve \giantfo\ goes as
\eqn\numbpa{
d \sim  n^2/2
}

In order to compute the charge under the $U(1)$ gauge field that performs translations
in $\psi$ we need know to
how many times this curve wraps the $CP^1$ inside $CP^2$ \refs{\BerensteinKE, \BarnesBW}. It is
easy to see that this number is $n$.
The amount of wrapping of this curve over the $CP^1$ in $CP^2$ is $n$. So the total
charge under the generator $J$ that rotates all the angles is
\eqn\totcha{
\hat q  =  \sum_i q_i= N n
}
We  define overall $U(1)$ charge $\hat q $ to be the change in phase when we shift $ \psi \to
\psi + 2 \pi$. So we have that $e^{i \hat q  \psi}$ is the shift in phase for a state of
charge $\hat q $.

Our strategy is as follows. The total charge that we have at our disposal is
$\hat q = 3 q$.
We split it   as $ \hat q = (3q - n N) + nN$. The second term will be
realized by $n$ giant gravitons and the first by momentum along $\psi$.
In other words, the $n$ giant gravitons are D3 branes that are intersecting at points
on the $CP^2$ (and form a smooth surface \giantfo ) and are coincident along the fiber parametrized
by $\psi$. We have many moduli of this configurations counted by \numbpa .
The momentum $L_0 \equiv (3q - n N)$ will be carried by these oscillations.
In other words the D3 branes wrapped along $\psi$ give us an effective string with
central charge $c = 6 d = 3 n^2$.

Then the entropy is
\eqn\entrop{
 S = 2 \pi \sqrt{ c L_0/6} = 2 \pi \sqrt{ { 1 \over 2} n^2 ( 3q - n N ) }
 }
 Note that we have not said anything about the value of $n$.  We now
 maximize the entropy with respect to $n$ we get
 \eqn\valn{
 n = { 2 q/N}
 }
 and substituting in \entrop\ we obtain
 \eqn\finres{
 S = 2 \pi \sqrt{2} { q^{3/2} \over N}
}
in agreement with \ccelow.

Notice, however that there is an important difference between the computation that
is done here and the usual computation for the $D1D5p$ system. In the latter case it is
possible to vary the parameters of the compactification to go to a regime where the
  amount of energy contained in momentum
is much smaller than the energy  of the D-branes, which is a
necessary condition for being able to view the momentum as producing
small oscillations on the D-branes. In the discussion of this
section it is not possible to satisfy this   condition. Equation
\valn\ implies that the energy contained in oscillations of the
branes is comparable to the brane tensions, and there is no obvious
parameter that we can adjust to change this fact\foot{ One would
like to increase the radius of the $\psi$ circle without changing
anything else, but this would not be a solution to the gravity
equations.}. As a consequence the discussion of this section falls
short of qualifying as a completely satisfactory derivation of
\finres\ (note, nonetheless, that all factors work bang on).

This point of view lets us also heuristically understand why we need to have angular
momentum $j_1$. This arises as follows. The system we had above is very similar
to the $D1D5p$ system in the NS sector, since
the fermions are anti-periodic in the $\psi$ direction.
Recall that the $D1D5p$ black hole
has $j_1=0$ \ref\BreckenridgeIS{
  J.~C.~Breckenridge, R.~C.~Myers, A.~W.~Peet and C.~Vafa,
  ``D-branes and spinning black holes,''
  Phys.\ Lett.\ B {\bf 391}, 93 (1997)
  [arXiv:hep-th/9602065].
}
(though $j_2$ can be non-zero\foot{
It can be seen that for small black holes the formulas in \cvetic\
allow $j_2$ to be non-zero with a bound similar to the one in \BreckenridgeIS .})
and this black hole naturally arises in the
Ramond sector. When we perform a spectral flow to the NS sector we get
$j_1 = {c \over 12}$. In our case, we cannot choose the fermions to be periodic along $\psi$ due
to the way the circle is fibered over $CP^2$. However, writing down the same formula
as we had for the $D1D5p$ in the NS sector we would get
 $j_1 = {c \over 12} =  {n^2 \over 4} = { q^2 \over N^2} $. On the other hand we get
 $j_1 = 3 {q^2 \over N^2}$ from
\ccelow , which has a different numerical coefficient. It would be nice to compute
$j_1$ properly and see whether it agrees with the black hole answer.

\newsec{Partition Functions over $\half$, ${1\over 4}^{th}$ and ${1\over
8}^{th}$ BPS States}
In this section  we will study the partition function over ${1 \over
8}^{th}$, a quarter and half BPS supersymmetric states in $\CN=4$
Yang Mills. We will compute these partition functions in free Yang
Mills, at weak coupling using naive classical formulas,  and at
strong coupling using the AdS/CFT correspondence. In the case of
quarter and ${1\over 8}^{th}$ BPS states, our free and weak coupling
partition functions are discontinuously different. However the weak
coupling and strong coupling partition functions agree with each
other (see \cachazo\ for an explanation).

It is possible that something similar will turn out to be true for
the ${1 \over 16}^{th}$ cohomology (see \wittenindex\ for a possible
mechanism). This  makes the enumeration of the weakly coupled $Q$
cohomology an important problem. We hope to return to this problem
in the near future.

 \subsec{Enumeration of ${1 \over 8}^{th}$, quarter and half BPS
Cohomology}
In this subsection we will enumerate operators in the anti-chiral ring,
i.e. operators that are annihilated by $  Q^{ \alpha 1}$,
with $ \alpha =\pm \half$, and their Hermitian conjugates (these are the
charges we called $Q$ and $Q'$ in previous sections \foot{If we had chosen states annihilated
by $ \bar  Q^{\dot \alpha}_1$ we would have obtained the chiral ring. }).
All such states have $\Delta=0$ and $j_1=0$. It is not
possible to isolate the contribution of these states to ${\cal I}_{YM}$
(note the index lacks a chemical potential that couples to $j_1$); nonetheless
we will be able to use dynamical information to count these states below.

This enumeration is easily performed in the free theory. Only the letters
$X, Y, Z, \psi_{0, \pm, +++}$ (see Table 2) and no derivatives
 contribute in this limit.  We will denote these letters by $ \bar \Phi_i$
($i=1 \ldots 3$) and ${\bar W}_{\dot \alpha}$ ($\dot \alpha= \pm$)
below. Note that these letters all  have $j_1=0$ and
$E=q_1+q_2+q_3$. The partition function \eqn\pfcom{Z_{cr-free}= Tr
\exp \left[ \sum_i \mu_i q_i + 2 \zeta j_2 \right]} is given by the
expression on the RHS of \pff\ with
\eqn\littlefdef{f_B=\sum_{i=1}^3e^{\mu_i}, ~~~f_F= 2 \cosh \zeta
e^{\mu_1+\mu_2+\mu_3 \over 2}.} Note that $1-f_B-f_F$ is positive at
small $\mu_i$ but negative at large $\mu_i$. We conclude that the
partition function \pfcom\ undergoes the phase transition described
in \refs{\sundborg, \sphere} at finite values of the chemical
potentials, and that its logarithm evaluates to an expression of
order ${N^2}$ at small $\mu_i$.

We now turn to the weakly interacting theory.  As explained in
\refs{\oferun,\cachazo}, at any nonzero coupling no matter how
small, the set of supersymmetric operators is given by all gauge
invariant anti-chiral fields $ \bar \Phi_i, \bar W_{\dot \alpha}$
modulo relations $[ \bar \Phi_i, \bar \Phi_j]= [\bar \Phi_i , {\bar
W}_{\dot \alpha} ] =0$ and $\{{\bar W}_{\dot \alpha}, {\bar W}_{\dot
\beta} \} =0 $ (the first of these follows from $\partial_{\bar
\Ph_i} {\bar W} =0$ where ${\bar W}$ is the superpotential). In
general there can be corrections to these relations (see \cachazo ).
We assume that such corrections do not change the number of elements in
the ring. In fact, if we go to the Coulomb branch of ${\CN = 4}$ we get a
$U(1)^N$ theory with no corrections at the level of the two derivative
action. The chiral primary operators at a generic point of this moduli space
are the same as all  the operators that we are going to count.

It is now easy to enumerate the states in the chiral ring.
The relations in the previous paragraph ensure that all the basic letters
commute or anticommute,  and so may be simultaneously  diagonalized, so we must
enumerate all distinct polynomials of traces of diagonal letters.
This is the same thing as enumerating all polynomials of the $3N$
bosonic and $2N$ fermionic eigenvalues that are invariant under
the permutation group $S_N$. We may now formally substitute the eigenvalues
$\bar \phi_i^f$ and $\bar W_{\dot \alpha}^f$ ($f=1 \ldots N$) with bosonic and fermionic
creation operators $a^f_i$ and $w^f_{\dot \alpha}$; upon acting on the
vacuum these produce states in the Hilbert space of $N$ particles,
each of which propagates in the potential of a 3 dimensional bosonic
and a 2 dimensional fermionic oscillator. The permutation symmetry
ensures that the particles are identical bosons or fermions depending on
how many fermionic oscillators are excited.  As a consequence
we conclude that the cohomological partition function is given by  the
coefficient of $p^N$ in
\eqn\parteigth{\eqalign{
Z_{1/8}(p,\gamma_1, \gamma_2,\gamma_3,\zeta) & =
\sum_{N=0}^\infty p^N Z_N(\gamma_1,\gamma_2,\gamma_3, \zeta) =
\cr
&=
 \prod_{n, m, r=0}^\infty
 { \prod_{s=\pm 1} (1 + p \, e^{ s \zeta} e^{
  -  (2 n +1) \gamma_1- ( 2 m +1) \gamma_2 - (2 r +1) \gamma_3 })
 \over (1 - p \, e^{ -  n 2 \gamma_1- m 2 \gamma_2 - r 2 \gamma_3 } ) ( 1 - p \,
 e^{
-  (2 n +2) \gamma_1-  (2m+2) \gamma_2 - (2 m +2)  \gamma_3 } )
  }
 }}

Further discussion on these 1/8 BPS states can be found in
\BerensteinAA .

We may now, specialize both the free and the interacting
cohomologies listed above to ${1 \over 4}^{{th}}$ BPS cohomology by
taking the limit $\gamma_3 \to \infty$. The only letters that
contribute in this limit are $\bar \Phi_1$ and $\bar \Phi_2$ ($X, Y$
of Table 2). The final result for the interacting cohomology may be
written as \eqn\partqr{ Z_{1/4}(p,\gamma_1, \gamma_2) =
\sum_{N=0}^\infty p^N Z_N(\gamma_1,\gamma_2) = { 1 \over \prod_{n, m
=0}^\infty
 (1 - p \, e^{ -   n 2 \gamma_1 -   m 2 \gamma_2} )  } }
 For a more explicit construction of 1/4 BPS operators see
\ref\D'HokerVF{
  E.~D'Hoker, P.~Heslop, P.~Howe and A.~V.~Ryzhov,
  ``Systematics of quarter BPS operators in N = 4 SYM,''
  JHEP {\bf 0304}, 038 (2003)
  [arXiv:hep-th/0301104].
} and references therein.

It is instructive to compare the $\gamma_3 \to \infty$ limit
\partqr\ of \parteigth\ to the same limit of the
partition function over $Q$ cohomology of the previous section that
also simplifies dramatically in this limit.The only letters that
contribute in this limit are $X,Y, \Psi_{+,++-} $ (where the indices
refer to $j_1, q_1, q_2, q_3$ charges). Further, it is easy to
verify that $ Q \Psi_{+, ++-} \propto [X,Y]$. As a consequence the
matrices $X$ and $Y$ should commute and may be diagonalized;
furthermore the matrix $\psi$ must also be diagonal (so that $Q$ anihillates it).
The cohomology in this
limit is thus given by the partition function of $N$ particles in a
2 bosonic and one fermionic dimensional harmonic oscillator.
\eqn\partcom{ Z = \sum_N p^N Tr[ y^{2 J_1} e^{ - \gamma_i L_i } ] =
\prod_{n,m \geq 0 } { (1 + p y e^{ - 2 (n+1)  \gamma_1 - 2 (m+1)
\gamma_2 } ) \over (1 - p e^{- 2 n \gamma_1  - 2 m \gamma_2} ) } }
The Index ${\cal I}^{WL}$ over this cohomology is then computed by setting
$y=-1$. At this special value, terms in the numerator with values
$m, n$ cancel against terms in the denominator with $m+1, n+1$
leaving only \eqn\idextwo{ {\cal I}_{YM}^{WL} = \sum_N
{\cal I}^{WL}_{YM}(N)= \sum_N p^NTr_N[ (-1)^{F} e^{ - \gamma_i L_i } ] =
 {  1  \over (1-p) \prod_{n=1}^\infty (1 -
 p e^{- n  2 \gamma_1})(1- p e^{- n 2 \gamma_2} ) }
 }
This is an exact formula for the $\gamma_3 \to \infty$ limit of the
index ${\cal I}^{WL}_{YM}$.  Multiplying it with $(1-p)$ and setting $p$ to
unity, we recover the large $N$ result \ymindex\ (at $\gamma_3 = \infty$).

It is also possible to further specialize \partqr\ to the half BPS
cohomology (of states annihilated by supercharges with $q_1 = \half$)
 by taking the further limit
$\gamma_2 \to \infty$ to obtain
\eqn\parthalf{
Z_{1/2}(p,\gamma_1) = \sum_{N=0}^\infty p^N Z_N(\gamma_1) =
{ 1 \over \prod_{n=0}^\infty
 (1 - p \, e^{ - n 2 \gamma_1} ) }
 }

Note that the free half BPS cohomology, interacting half BPS
cohomology and the $\gamma_2, \gamma_3 \to \infty$ limit of
${\cal I}^{WL}_{YM}$ all coincide. On the other hand the free quarter BPS
cohomology sees many more states than the interacting quarter BPS
cohomology which, in turn, sees a larger effective number of states
than the $\gamma_3 \to \infty$ limit of the index. The last
quantity, the index, receives contributions from $\bar \Phi_1,~ \bar
\Phi_2$ and  $\psi_{+,0;++-}$, which are all the states in table two
which have $L_3 =0$. This index  sees a smaller number of states as
a consequence of cancellations involving the presence of the fermion
$\psi_{+,0;++-}$. Again, the ${1 \over 8}^{th}$ BPS free cohomology
sees more states than the interacting cohomology, which in turn sees
more states than the index, with no restrictions on chemical
potentials. More explicitly, we can see that for very large charges,
or very small chemical potentials the entropy of \parthalf\ is that
of $N$ harmonic oscillators, which correspond basically to the
eigenvalues. Similarly, \partqr, and \parteigth\ give the entropy of
$2N$ and $3N$ harmonic oscillators respectively. All these entropies
are basically linear in $N$ in the large $N$ limit. The intuitive
reason is that the matrices commute, and so do not taking advantage
of the full non-abelian structure of the theory.

\subsec{ Protected double trace operator in the ${\bf 20}$}

As an example of a practical application of the exact partition
function over the chiral ring (derived in the previous subsection)
and the index ${\cal I}_{YM}$ (defined and computed in sections 3
and 4), in this subsection we will demonstrate that the scaling
dimension of a particular double trace operator Yang Mills operator
is protected.\foot{
This subsection was added in the revised version to answer a question raised by
M. Bianchi.}

Consider  $SU(N)$ $\CN=4$ Yang Mills theory. Let us first study the
set of operators with quantum numbers $(q_1, q_2, q_3)=(3,1,1)$ and
$j_1=j_2=0$. Using $\Delta \geq 0$ we conclude that such operators
have $E \geq 5$; we will be interested in operators that saturate
this equality. Let us first consider free Yang Mills theory. The set
of all such operators is easy to list; we find
\eqn\opercontr{\eqalign{ & Tr[W_\alpha W^\alpha] Tr[X^2] ~,~~~~~Tr[
W_\alpha X] Tr[W^\alpha X] \cr &Tr[X^2]Tr[XYZ] , ~~~Tr[X^2]Tr[XZY]
~,~~Tr[XY]Tr[X^2 Z] ~,~~~~~Tr[XZ]Tr[X^2 Y] \cr & Tr[YZ] Tr[X^3] }}
Turning now to the interacting theory, we note that all but one of
these operators belongs to the 1/8 BPS chiral ring, and so has
protected scaling dimension \foot{In general the interacting
operator with good scaling dimension will have a complicated form,
admitting admixtures with single trace operators.}. Indeed it is not
difficult to check that the appropriate coefficient in
\parteigth\ (after subtracting the $U(1)$ part and the single trace
contribution) is 6 implying that 6 of the 7 operators in \opercontr\
have protected dimensions. The unprotected operator in \opercontr\
is simply $O'=tr[X^2] tr[X[Y,Z]]$.

Note that the operators studied in the previous paragraph have
$\Delta=0$, $L_1=6$, $L_2=2$, $L_3=2$, $J_2=0$. Notice that states
with quantum numbers $(q_1, q_2, q_3)=(5/2, 1/2, 1/2)$, $j_1=\half$,
$j_2=0$ share these values for $\Delta, L_i, j_2$; (and, moreover,
are unique in this regard in the double trace sector of the $SU(N)$
theory). As a consequence we will now list all double trace
operators in the free theory with these quantum numbers. They are
\eqn\tradoub{\eqalign{ & tr[ \psi_{+,-++} X] tr[X^2] ,
~~~tr[\psi_{+,+-+} Y] tr[X^2] ,~~~~~
 tr[\psi_{+,++-} Z] tr[X^2]\cr
& tr[\psi_{+,+-+}X] tr[XY]
 ~,~~~~~~
 tr[\psi_{+,++-} X] tr[XZ] ~,~~~~
}}

It follows from the discussion above that the contribution of
double trace operators with $L_1=6, L_2=L_3=4, J_2=0$ to the index
${\cal I}_{YM}$ is $(7-5)e^{-6 \gamma_1-2\gamma_2 -2\gamma_3} =
2e^{-6 \gamma_1-2\gamma_2 -2\gamma_3}$. As ${\cal I_{YM}}$ is not
renormalized, it must be that 4 out of the 5 operators listed in
\tradoub\ are exactly protected. The single non protected operator
is easily identified; at infinite $N$ it is the operator
\eqn\npdt{ O=\left( tr[ \psi_{+,-++} X]+ tr[\psi_{+,+-+} Y]  +
tr[\psi_{+,++-} Z] \right) tr[X^2]=Q_{+ \half, 1} tr [X {\bar X}
+Y{\bar Y} + Z {\bar Z}] tr X^2}   Note that $Q_{- \half, 1} O
\propto O'$; we see that the the two non protected operators $O$
and $O'$ are married together in the same long multiplet.

We have concluded that four double trace operators of the form
\tradoub\ are exactly protected. At the end of this subsection  we
will demonstrate that while three of these four operators are
$SU(1,2|3)$ descendants, a fourth is and $SU(1,2|3)$ primary. As
we have explained in section 3, the decomposition of the index
${\cal I}_{YM}$ into characters of $SU(1,2|3)$ yields information
about linear combinations of the number of short representations
of the Yang Mills theory. In the case at hand, the existence of
precisely one protected primary with these quantum numbers implies
the existence of exactly one double trace $cc$ type representation
with quantum numbers $q_1=2, q_2=q_3=0$, (or $R_1=R_3=0, R_2=2$) ,
$j_1=j_2=0$ (such a representation has $E=4$). This is an operator
of the schematic form $ O_{IJ} Q_{JK} - { 1\over 6} \delta_{IK}
O_{LJ} Q_{JL} $ where $O_{IJ}$ is the single trace operator in the
${\bf 20}$ of $SO(6)$. (This form is schematic because this
operator will mix with single trace operators, see for example
\bianchimixing ). Indeed this operator was studied in
\refs{\frolovone,\rossi,\edennew} and a proof that it is protected
was
  was given in \refs{\eden,\excep,\frolovtwo},  based on the
non-renormalization theorem in \petkou. The arguments of this
subsection may be regarded as an alternate proof of this non
renormalization.

To complete this subsection we will now demonstrate that 3 of the
operators in \tradoub\ are $SU(1,2|3)$ descendants. In fact the
operators in question will turn out to be descendants of 1/2 and
1/4 BPS states in $bb$ representations. In other words, some of
them result from the action of $SU(1,2|3)$ generators on conformal
primaries which have lower conformal weight.  So let us understand
the protected $bb$ representations with $E=4$.
 One of them arises from  the 1/2 BPS chiral primary operator
$tr[X^2]^2$. We can now consider the $SU(1,2|3)$ descendants of it.
By analyzing in more detail the action of the supercharges we find
that two of the states in \tradoub\ are $SU(1,2|3)$ descendants of
$tr[X^2]^2$. Another operator that we should consider is the 1/4 BPS
double trace operator that is in the ${\bf 84}$ (the protected
nature of this operator follows from the partition function of
chiral primary operators \parteigth , or \partqr -in other words, it
gives rise to operators in the chiral ring).  This operator has
$SU(4)$ Dynkin labels $R = (2,0,2)$. It turns out that there is one
$SU(1,2|3)$ descendant of the ${\bf 84}$ with the quantum numbers
appearing in \tradoub, completing our demonstration.

\subsec{Large $N$ limits and Phase Transitions}
In this subsection we will study the large $N$ limit of the
partition functions \parteigth, \partqr, \parthalf. We will first
briefly consider the limit $N \to \infty$ at fixed values of the
chemical potentials, and show that in this limit these partition
functions reproduce the supergravity answers \oneeigthgamma . We will then
turn to large $N$ limits in which the chemical potentials scale with
$N$. We find that the formulas for 1/4 and 1/8 BPS states lead to
large $N$ phase transitions. This phase transition is the
Bose-Einstein condensation of the lowest mode, the ground state of
the harmonic oscillators we had in the previous subsection.

In the $N\to \infty$ and fixed chemical potential the partition
functions \parthalf, \partqr, \parteigth , become independent of
$N$. This limit is most easily evaluated by multiplying the
partition functions by $(1-p)$ \foot{This step cancels the divergent
contribution of the ground state of the harmonic oscillator in this
limit. We will have a lot more to say about this below.} and setting
$p=1$. The entropy then grows as a gas of massless particles in one,
two and three dimensions respectively.

For half BPS states we have
\ref\CorleyZK{
  S.~Corley, A.~Jevicki and S.~Ramgoolam,
  ``Exact correlators of giant gravitons from dual N = 4 SYM theory,''
  Adv.\ Theor.\ Math.\ Phys.\  {\bf 5}, 809 (2002)
  [arXiv:hep-th/0111222].
}
 \eqn\parthalfln{ Z_{1/2}(\gamma_1) = { 1
\over \prod_{n=0}^\infty
 (1 - \, e^{ - n 2 \gamma_1} ) }
 }
Clearly, in the large $N$ limit,
 \parthalfln\ may be thought of as the multiparticle
partition function for a system of bosons with
\eqn\zsphalf{Z_{{1/2}-sp}=\sum_{n=1}^\infty e^{-2 n
\gamma_1}={1\over 1-e^{-2 \gamma_1}} -1 ; } note that \zsphalf\ is
the same as the supergravity result \oneeigthgamma\ in the limits
 $\gamma_2 \to \infty$, $\gamma_1 \to \infty$. Similarly the large
$N$ limit of
\partqr\ is the multiparticle partition function for a system of
bosons whose single particle partition function is
\eqn\zspquater{Z_{{1/4}-sp}=\sum_{n,m} e^{-2 n \gamma_1 -2 m \gamma_2
}={1\over (1-e^{-2 \gamma_1}) (1-e^{-2 \gamma_2}) } -1 ,} which is
the same as \oneeigthgamma\ in the limit $\gamma_3 \to \infty$. In a
similar fashion, in the large $N$ limit of
\parteigth\ is precisely the multiparticle partition function \parteigthstrong\  a
system of bosons and fermions whose single particle partition
functions are given by \oneeigthgamma.

We now turn to large $N$ limits of these partition functions in
which we will allow the chemical potentials to scale with $N$. Let
us start with the $1/2 $ BPS case, and set $\gamma_1=\gamma$. This
case does not have a phase transition.
 We write
 \eqn\integ{
 \log Z(\gamma,p) = -\sum_n \log(1 - p \, e^{-2  n \gamma} ) \sim - { 1 \over 2 \gamma }
 \int_0^\infty dx \log(1-p \, e^{-x} )
 }
We can first solve for $p$ by writing \eqn\solp{ N = p \,
\partial_p \log Z = { 1\over 2 \gamma } \int_0^\infty dx
 { p \, e^{-x} \over 1- p\, e^{-x} } =
- { 1 \over 2 \gamma } \log (1-p)
}
We can now write $\tilde \beta \equiv N 2 \gamma $. Then \solp\ is independent of $N$ and
it has a solution for all values of $\tilde \beta$ . We can then write the partition
function as
\eqn\integsc{
\log Z_N(\gamma) = N \left\{ { 1\over \tilde \beta} \int_0^\infty dx
\log[ 1 - (1 - e^{ - \tilde \beta} ) e^{- x} ] -  \log (1 - e^{-\tilde \beta} )  \right\}
}
We see that this formula is of order $N$. There is no transition as a function of
$\tilde \beta$. For large  values of $\tilde \beta \ll 1$,
 it turns out
that \integsc\ is independent of $N$ when expressed in terms of
$\gamma$. This can be most easily seen by setting $p=1$ in \integ .
As expected the change in behavior happens at a temperature $(2
\gamma)^{-1} \sim N$ which is when the trace relations start being
important. For very small $\tilde \beta$ we find that \integsc\
becomes $\log Z_N \sim N[ - \log \tilde \beta +1] $, which captures
the large temperature behavior of $N$ harmonic oscillators plus an
$1/N!$ statistical factor.

Let us now consider 1/4 BPS states. Let us set $\gamma_1 = \gamma_2 = \gamma$.
 For sufficiently large temperatures we approximate
the partition function as
\eqn\integqu{
 \log Z(\beta,p) = \sum_{n_1, n_2} - \log(1 - p e^{- (n_1 + n_2) 2 \gamma } )
 \sim { 1 \over (2 \gamma)^2}
 \int_0^\infty dx x [- \log(1-p e^{-x} )]
 }
Now we find a new feature when we compute
\eqn\compn{
N = { 1 \over (2 \gamma)^2 } \int dx x{ p e^{-x} \over 1- p e^{-x} } = { 1\over (2 \gamma)^2}
Pl[2, p]
}
where $Pl[2,p]$ is the PolyLog function.
Now we see that for the lowest value of the chemical potential, $p=1$,
 we get
\eqn\limnmax{
N_{max} = { 1 \over (2 \gamma)^2} { \pi^2 \over 6}
}
Defining $\tilde \beta \equiv  2 \gamma \sqrt{N}$
we see that there is a critical temperature, ${\tilde \beta}_c^2 = {\pi^2 \over 6}$,
at which there is a phase transition obtained by setting $N_{max} = N$ in \limnmax .
 At temperatures smaller than this temperature we have a Bose-Einstein condensation of
 the ground state of the harmonic oscillator. In this regime the free energy $Z_N(\beta)$
 is given by
 \integqu\ with $p=1$. For higher temperatures we are supposed to solve for $p$ using
 \compn\ and then insert it in \integqu\ to compute the free energy.
 We get
 \eqn\freennew{
 \log Z_N(\tilde \beta') = N \left\{{ 1 \over {\tilde {\beta}}^2}
 \int_0^\infty dx x [- \log(1-p(\tilde \beta)  e^{-x} )]  - \log p(\tilde \beta)
\right\} } where $p(\tilde \beta)$ is the solution to \compn .
The for large temperatures we have $\log Z_N \sim N[ - \log \tilde \beta^2 +1]$ which
captures the entropy of $N$ 2-dimensional harmonic oscillators plus the $1/N!$ statistical
factor.
It is possible to see that at $\tilde \beta_c$ we have a second order phase transition.

One can find similar results for the 1/8 BPS states. We set $\gamma_i = \gamma$.
 In this case
the rescaled temperature is given by $\tilde \beta' = 2\gamma
N^{1/3}$. The results are similar. For low temperatures the answer
is independent of $N$ and for high temperatures we have a free
energy which is linear in $N$ and is a function of the rescaled
temperature $\tilde \beta'$. Again there is a second order phase
transition corresponding to the Bose-Einstein condensation of
ground state of the harmonic oscillator. If we think of these
harmonic oscillators as arising from D3 branes wrapping the $S^3$,
then we could think of this condensation as responsible for the
fact that the $S^3$ is contractible, in the spirit of the
transition in \ref\GreeneHU{
  B.~R.~Greene, D.~R.~Morrison and A.~Strominger,
  ``Black hole condensation and the unification of string vacua,''
  Nucl.\ Phys.\ B {\bf 451}, 109 (1995)
  [arXiv:hep-th/9504145].
}. It would be nice to see if this can be made more precise.

\newsec{Discussion}
In this paper we have considered an index that counts protected
multiplets for general four dimensional superconformal field
theories. This quantity captures all the information about
protected multiplets that can be obtained purely from group theory.

We have focused on the applications of this index to the $\CN=4$
Yang Mills theory. It is possible that (and would be interesting if)
our index turns out to be a useful tool in the study of $\CN=1$ and
$\CN=2$ superconformal theories as well.

Indices of the form that we have constructed have obvious
counterparts in superconformal theories in $d=3, 5, $ and $6$. It is
possible that some of these indices (whose theory we have not worked
out in detail) could have interesting applications.

The later half of this paper was devoted to a study of the
supersymmetric states of $\CN=4$ Yang Mills on $S^3$. We computed
this index for this theory and found that it precisely agrees with the
free supergravity spectrum when we take the large $N$ limit. The
index, however, does not reflect the large entropy of BPS black
holes in $AdS_5$. This is not a contradiction because there is no
clear argument from the supergravity point of view which says that
the black holes should contribute to the index.

A satisfactory Yang Mills accounting of the entropy of the BPS black
holes of \refs{\reallone,\reallthree, \cvetic} remains an important
outstanding problem. We have not even aware of a field theoretic
understanding of a rather gross feature of these black holes; the
fact that supersymmetric solutions are known only when certain
special relation between the charges is obeyed.

We think it should be possible  to use weakly coupled Yang
Mills theory to count the entropy of BPS black holes in $AdS_5\times
S^5$. In such a counting
 one will have to put in
some extra information about the dynamics of the theory (over and
above the superconformal algebra), see section 6. In this connection
it is encouraging that the counting of BPS states in the free theory
(without the $(-1)^F$) has some qualitative agreement with the black
hole results.
\bigskip

\centerline{\bf Acknowledgments}

We would like to thank O. Aharony, D. Berenstein, M. Bianchi, R.
Gopakumar, L. Grant, S. Lahiri, J. Marsano, K. Papadodimas, H.
Reall, R. Roiban, G. Rossi, N. Seiberg, A. Strominger and    M.
Van Raamsdonk   for useful discussions. The work of JM was
supported in part by DOE grant \#DE-FG02-90ER40542. The work of SM
and SR was supported in part by an NSF Career Grant PHY-0239626,
DOE grant DE-FG01-91ER40654, and a Sloan Fellowship.

\vfil

\appendix{A}{The $d=4$ Superconformal Algebra}

\subsec{The Commutation Relations} \eqn\allcommut{ \eqalign{
[(J_1)^{\alpha}_{\beta}, (J_1)^{\gamma}_{\delta}] &=
\delta^{\gamma}_{\beta} (J_1)^{\alpha}_{\delta} -
\delta^{\alpha}_{\delta} (J_1)^{\gamma}_{\beta} \cr
[(J_2)^{\dot{\alpha}}_{\dot{\beta}},
(J_2)^{\dot{\gamma}}_{\dot{\delta}}] &=
\delta^{\dot{\gamma}}_{\dot{\beta}}
(J_2)^{\dot{\alpha}}_{\dot{\delta}} -
\delta^{\dot{\alpha}}_{\dot{\delta}}
(J_2)^{\dot{\gamma}}_{\dot{\beta}} \cr
[(J_1)^{\alpha}_{\beta},P^{\gamma \dot{\delta}}] &=
\delta^{\gamma}_{\beta} P^{\alpha \dot{\delta}} - (1/2)
\delta^{\alpha}_{\beta} P^{\gamma \dot{\delta}}\cr
[(J_1)^{\alpha}_{\beta},K_{\gamma \dot{\delta}}] &=
\delta_{\gamma}^{\alpha} K_{\beta \dot{\delta}} - (1/2)
\delta^{\alpha}_{\beta} K_{\gamma \dot{\delta}}\cr
[(J_2)^{\dot{\alpha}}_{\dot{\beta}},P^{\dot{\delta} \gamma}] &=
\delta^{\dot{\delta}}_{\dot{\beta}} P^{\dot{\alpha} \gamma} - (1/2)
\delta^{\dot{\alpha}}_{\dot{\beta}} P^{\dot{\delta} \gamma}\cr
[(J_2)^{\dot{\alpha}}_{\dot{\beta}},K_{\dot{\delta} \gamma}] &=
\delta_{\dot{\delta}}^{\dot{\alpha}} K_{\dot{\beta} \gamma} - (1/2)
\delta^{\dot{\alpha}}_{\dot{\beta}} K_{\dot{\delta} \gamma}\cr
[H,P^{\alpha \dot{\beta}}] &= P^{\alpha \dot{\beta}} \cr
[H,K^{\alpha \dot{\beta}}] &= -K^{\alpha \dot{\beta}} \cr [K_{\alpha
\dot{\beta}}, P^{\gamma \dot{\delta}}] &=
\delta^{\dot{\delta}}_{\dot{\beta}} (J_1)^{\gamma}_{\alpha}+
\delta^{\gamma}_{\alpha}(J_2)^{\dot{\delta}}_{\dot{\beta}} +
\delta^{\dot{\delta}}_{\dot{\beta}} \delta^{\gamma}_{\alpha} H \cr
[(J_1)^{\alpha}_{\beta},Q^{\gamma n}] &=  \delta^{\gamma}_{\beta}
Q^{\alpha n} - (1/2) \delta^{\alpha}_{\beta} Q^{\gamma n} \cr
[(J_2)^{\alpha}_{\beta},\bar{Q}^{\gamma}_{ n}] &=
\delta^{\gamma}_{\beta} \bar{Q}^{\alpha }_{n} - (1/2)
\delta^{\alpha}_{\beta} \bar{Q}^{\gamma}_{n} \cr [K_{\alpha
\dot{\beta}},Q^{\gamma n}] &= \delta^{\gamma}_{\alpha}
\bar{S}^{n}_{\dot{\beta}} \cr [P_{\alpha
\dot{\beta}},\bar{Q}^{\gamma}_{n}] &= \delta^{\gamma}_{\alpha} S_{n
\dot{\beta}} \cr [H,Q^{\gamma n}] &= \half Q^{\gamma n} ,~~~~[H,
{\bar Q}^{{\dot \alpha}}_n]=\half {\bar Q}^{{\dot \alpha}}_n, ~~~
[H, S_{\alpha n}]= -\half S_{\alpha n}, ~~~ [H, {\bar S}_{{\dot
\alpha}}^n] =-\half {\bar S}_{{\dot \alpha}}^n \cr
 [r,Q^{\gamma n}] & = Q^{\gamma
n} ,~~~~[r, {\bar Q}^{{\dot \alpha}}_n]=-{\bar Q}^{{\dot
\alpha}}_n,~~~~[r,S_{\alpha i}] =- S_{\alpha i} , ~~~[r, {\bar
S}_{{\dot \alpha }}^i] =S_{{\dot \alpha }}^i \cr \{S_{\alpha i},
Q^{\beta j} \} &= \delta^{j}_{i} (J_1)^{\beta}_{\alpha} +
\delta^{\beta}_{\alpha} R^{j}_{i} + \delta^{j}_{i}
\delta^{\beta}_{\alpha} ({H \over 2} + r{4 - m \over 4 m}) \cr
\{Q^{\alpha m}, {\bar Q}^{\dot \alpha}_m \} &= P^{\alpha {\dot
\alpha}} \delta^n_m \cr \{ S_{\alpha m} , {\bar S}_{\dot \alpha}^m
\} &= K_{\alpha {\dot \alpha}} \delta^n_m \cr
 [R^{j}_{i}, Q^{\alpha p}] &=
\delta^{p}_{i} Q^{\alpha j} - (1/m) \delta^{j}_{i} Q^{\alpha p} \cr
[R^{m}_{n}, R^{p}_{q}] &= \delta^{m}_{q} R^{p}_{n} - \delta^{p}_{n}
R_{q}^{m} }}  The Cartan generators are ${\cal H}$,${\cal J}_1 =
(J_1)^{2}_{2} = -(J_1)^{1}_{1}$,  ${\cal J}_2 = (J_2)^{2}_{2} =
-(J_2)^{1}_{1}$, ${\cal R}_n = R^{n}_{n} - R^{n+1}_{n+1}$. While we
have used script letters here, the Cartan generators above are the
same as those used in the rest of the paper. The eigenvalue under
${\cal H}$ is the energy $E$, the eigenvalues under ${\cal J}_1,
{\cal J}_2$ are the angular momenta $j_1,j_2$ and the eigenvalues
under ${\cal R}_i$ are the R-charges $r_i$. Notice that in the way
that we have defined the generators the commutation relations of the
$J$s and the $R$s differ by a sign. For this reason, in the case of
$m=2$, the structure of BPS states and null vectors is not symmetric
under the exchange of the  $J$ and the $R$ quantum numbers.

\subsec{An Oscillator Construction of the Algebra}
It is possible to find an explicit
oscillator construction of this algebra following \BarsEP. We introduce two sets of bosonic oscillators $a^{\alpha}, b^{\dot{\alpha}}, \alpha, \dot{\alpha} = 1,2$
with adjoints $a_{\alpha},b_{\dot{\alpha}}$. In addition, we introduce fermionic
oscillators $\alpha^{n}$ with adjoints $\alpha_{n}, n = 1 .. 4$. As
expected the $a$ and $b$ oscillators will transform as Lorentz
spinors whereas the fermionic oscillators will transform in the
fundamental representation of $SU(4)$. The generators of the superconformal group are now defined as below:

\eqn\opdef{
\eqalign{
H &= \half (a^\alpha a_\alpha + b^{\dot{\alpha}} b_{\dot{\alpha}}) + 1 \cr
(J_1)^{\alpha}_{\beta} &= a^{\alpha} a_{\beta} - \half \delta^{\alpha}_{\beta} a^{\gamma} a_{\gamma} \cr
(J_2)^{\dot{\alpha}}_{\dot{\beta}} &= b^{\dot{\alpha}} b_{\dot{\beta}} - \half \delta^{\dot{\alpha}}_{\dot{\beta}} b^{\dot{\gamma}} b_{\dot{\gamma}} \cr
P^{\alpha \dot{\alpha}} &= a^{\alpha} b^{\dot{\alpha}} \cr
K_{\alpha \dot{\alpha}} &= a_{\alpha} b_{\dot{\alpha}} \cr
R^{n}_{s} &= \alpha_{s} \alpha^{n} - {1 \over m} \delta^n_s \alpha_{t} \alpha^{t} \cr
Q^{\alpha n} &=  a^{\alpha} \alpha^{n} \cr
\bar{Q}^{\dot{\alpha}}_{n} &=  b^{\dot{\alpha}} \alpha_{n} \cr
S_{\alpha n} &= a_{\alpha} \alpha_{n} \cr
\bar{S}_{\dot{\alpha}}^{n} &=  b_{\dot{\alpha}} \alpha^{n} \cr
r &= \alpha_n \alpha^n \cr
C &= b^{\dot{\alpha}} b_{\dot{\alpha}} - a^{\alpha}a_{\alpha}  - \alpha_{n} \alpha^{n}
}}
While $C$ appears in the oscillator construction it commutes with all the generators of the algebra and is not really part of it. When we construct representations
of the algebra  using oscillators we fix the total value of $C$.

\appendix{B}{Algebraic Details Concerned with the Index}

\subsec{Superconformal Indices from Joining Rules}
As we have explained in section 2, any index is given by the sum
\indexdef\ where the sum runs over representations of the algebra, and
the coefficients $\alpha[i]$ are chosen such
that $I$ evaluates to zero on every combination of short
representations that can pair into a long representation.

The simplest indices for the superconformal algebra are given by $\alpha[i]= 0$
for all $i \neq i_0$ for some specific $i_o$; this choice of $\alpha[i]$ defines an
index only when the representation $i_0$ never makes an appearance
on the right hand side of \charac. An inspection of \charac\ and
Table 2 shows that this is true of the representations of
the form ${\bf bx}$ with $R_1=0$ or $R_1=1$ and representations of the
form ${\bf xb}$ with any of $R_{m-1}=0$ or $R_{m-1}=1$. The number of all
such representations constitutes an index.

We briefly pause to list these special representations in the most
physically relevant cases $m=1,2, 4$. Protected representations do
not exist in the $\CN=1$ algebra $(m=1)$. In the $\CN=2$ algebra
$(m=2)$ they consist of $SU(2)_R$ singlets with $j_1=0$ and
$E=2j_2+r/2$, $SU(2)_R$ doublets with $j_1=0$ and $E=2j_2+r/2+2$ and
chirality flips ($j_1 \leftrightarrow j_2$, $r \leftrightarrow -r$)
of these. In the $\CN=4$ $(m=4)$ algebra they are representations
with $R_1=0=j_1$ and $E=2j_2+R_2+R_3/2$, with $R_1=1$, $j_1=0$ and
$E=2j_2+3/2+R_2+R_3/2$, and the chirality flips ($j_1
\leftrightarrow j_2$, $R_i \leftrightarrow R_{4-i}$) of these. Note
that this includes representations with $j_1=j_2=R_1=R_3=0$ and
$E=R_2$; these are the famous chiral primaries (gravitons) of the
$\CN=4$ theory.

Let us now turn to indices that have support on representations that
do appear on the RHS of \charac. We first consider indices built out
of representations of the form ${\bf \tilde{c}a}$. It follows from
the first equation in \charac\ that, on an index
\eqn\alphaeq{\alpha[{\bf \tilde{c}a}_{j_1, j_2, r, R_1, R_j}]+
\alpha[{\bf \tilde{c}a}_{j_1-\half, j_2, r+1, R_1+1, R_j}]=0 .} To
begin with let us assume $m>1$. Notice that the two representations
that appear in \alphaeq\ have equal values of
\eqn\qnsecond{j_2,~~~r' \equiv r-R_1~~~, M \equiv R_1+2j_1 , ~~~R_j
~~(j=2 \ldots m-1).} The number of representations with given values
for these conserved quantum number is $M+2$ (recall $j_1$ varies in
half integer units from $-\half $ upto $M/2$; see Table 2). The
$\alpha$ coefficients for these representations are constrained by
$M+1$ equations. We conclude that there is exactly one index for any
given set of charges \qnsecond\ that obeys $E_1>E_2$; this index is
given by \firstind.

In a very similar fashion, we conclude that \secondind\ also defines an index provided $E_2 > E_1$.

If $E_1=E_2$ the last equation in \charac\ applies. We
have a total of $(M+2)(N+2)$ representations with given values for
$M$, $N$, $r''' \equiv r+2 j_1 - 2 j_2$ and $R_l$. The $\alpha's$
corresponding to these representations are constrained by
$(M+1)(N+1)$ equations (from the last equation in \charac). This
leaves us with an $M+2+N+2 -1$ dimensional linear vector space of
Indices. A convenient basis for these Indices  is given by  $N+2$
indices \firstind\ (see the LHS of the equation below) ( plus the
$M+2$ indices of the form \secondind, (see the RHS of the equation
below) subject to the single additional constraint \consttrr.

Finally we turn to the special case $m=1$ (the $\CN=1$ algebra). As
we have remarked above, no representations are absolutely protected
in this case. The two indices \firstind\ and \secondind, formally
continue to be protected; however the expressions for these indices
\eqn\firstindo{\eqalign{
I^L_{j_2, r'} &= \sum_{p=-1}^{\infty} (-1)^{(p+1)}
n[{\bf \tilde{c}x}_{{p\over 2}, j_2, r'+p}] \cr
I^R_{j_1,r''} &= \sum_{p=-1}^{\infty} (-1)^p n[{\bf x\tilde{c}}_{j_1, {p\over
2}, r''-p}]}}
 (where $r'=r+2j_1$ and $r''=r-2j_2$) now involves a sum
over an infinite number of representations, and so could diverge.

\subsec{The Index ${\cal I}^{WL}$ as a sum over characters}
We define $Q' \equiv Q^{\half, 1} $, the $SU(2)$ partner of $Q \equiv
Q^{-\half, 1} $. $Q'$ has
 charge $\Delta=-2$. All other supercharges either have $\Delta =0$
($Q$ itself and   all the supercharges in the
$SU(1,2|m-1)$ subalgebra) or $\Delta=2$ (all other supercharges).
There are also negative $\Delta$ bosonic generators. For example the
$SU(2)$ spin operator $J^{+}_1$ has $\Delta = -2$ and it appears in the
anticommutation relation between $S$ and $Q'$. We also get negative
$\Delta$ states among the raising operators of $SU(m)$.

Notice that we can rederive some of the results in section two as follows.
We start with the anticommutation relation \qscom . From this we derive
that all states should obey $\Delta \geq 0$.
Now suppose that we start with the highest weight state with Cartan charges
$|\psi_0\rangle = |E, j_1, j_2,r, R_i \rangle$.
This state has the lowest value of
$\Delta$ among all the level zero states, which is $\Delta_0 = E-(E_1 - 2)$, where
$(E_1-2)$ is the combination of charges appearing in the right hand side of \qscom .
Notice that we cannot lower the value of $\Delta$ by acting with bosonic generators
since this is a highest weight state under the $SU(2)\times SU(2) \times SU(m)$ subalgebra.
 We now start acting with the supercharges
$Q^{\alpha , i}$ on this state. When we act with $Q'$ we lower the value of $\Delta$.
If $Q'$ does not annihilate the state we conclude that $\Delta_0 - 2 \geq 0$.
If it is strictly bigger, then we have the representations of the generic type, which
we called ${\bf a}$ in section two. When it becomes equal to zero, then we get
representations of the type ${\bf c}$, which obey $E = E_1$. The final possibility is
that $Q'$ annihilates the level zero state. Using the anticommutator of
$Q'$ and $S'$ we notice that this can happen only if $\Delta_0 + 4 j_1=0$.
Since, $\Delta_0 \geq 0$, this implies that $j_1=0$ and $E = E_1 -2$. So we have
a representation called  ${\bf b}$ in section two.

Using some of these ideas it is possible also to understand the structure of the null
vectors. For that it is convenient to consider level one  states of the form
 $J^- Q' |\psi_0\rangle$ and $Q |\psi_0 \rangle$. Using the anticommutation relations
 it is possible to show that the determinant of the $2 \times 2$ matrix of inner products
 among these states is proportional to $ 2j_1 (\Delta_0 -2)$. So we find that in the
 case that $\Delta_0=2$ we have a null state. It is also easy to show that
 the state $ Q'|\psi_0 \rangle$ has positive norm (if $\Delta_0 \geq 2$).
 This latter state transforms
 in the representation with highest weights $(E+\half, j_1 + \half , j_2 , r+1,R_1+1, R_j)$.
 Thus the zero norm state we just mentioned transforms in the representation of
 the form $(E+\half ,j_1- \half  , j_2 , r+1, R_1+1, R_j )$. This follows from the fact that
 we have one state with these quantum numbers plus the fact that if we had any other
 states with higher weights then we would be able to decrease the value of
 $\Delta$ for this state below $2$ and we could change it
 in a continuous fashion (as we increase the energy
 of the original state away from the value that makes it a ${\bf c}$ type representation), but
 this is not possible. So we have recovered the statements in section two about the
 structure of the null vectors of the ${\bf c}$ type representations, at least for the
 case $j_1 >0$. We can similarly continue the analysis for the structure of the null
 vectors of the ${\bf c}$ representations with $j_1=0$.

We can use some of these facts also to learn about the structure of the
states that contribute to the index.
For this purpose we should note that we get states with $\Delta=0$ by applying
$Q'$ to $|\psi_0\rangle$. This state has Cartan charges
\eqn\cartch{
(E+\half, j_1 + \half, j_2,r+1,R_1+1,R_i)
}
It is also easy to see that
it
transforms in the $SU(1,2|m-1)$ representation with charges $(E',j_2,r',R_j)$ given
in terms of the map \qnfirst\ applied to \cartch .
In this way we obtain formula \indexchar , for type ${\bf c}$ BPS representations.
The factor $(-1)^{2j_1 +1}$ comes from the statistics factor associated to
\cartch\ which will not be included when we consider the character in the
subalgebra, whose sign depends only on the $j_2$ quantum numbers.

For type ${\bf b}$ BPS representations the highest weight state itself has $\Delta =0$,
so we find that $(E',j_2',r',R_j)$ are directly given by the formula in \qnfirst\ in
terms of the Cartan eigenvalues of $|\psi_0\rangle $ and we do not get any overall
minus sign. This is summarized in \defqbc .

The character on the $SU(1,2|m-1)$
 character  in  \indexchar\ manifestly depends on the quantum numbers
$j_1, r, R_1$ of a ${\bf \tilde c}$
representation only through the combination $j_1 + r/2$, $r-R_1$.
This leads to \indconc\ and \fineq\ and \defqbc\ for ${\bf \tilde c}$ representations.
\subsec{Representation Theory of the Subalgebra $SU(2,1|m-1)$ }
The representation theory of the subalgebra is easily worked out, and
closely mimics the pattern presented in the previous section.
Briefly, representations are labeled by  the quantum numbers $(E',
j_2', r', R_i')$  that specify the $U(1) \times SU(2) \times U(m-1)$
($i= 2 \ldots m-1$) quantum numbers of the lowest weight state. Acting on
this lowest weight state with the supersymmetries charged under $J'_2$,
we find a set of level one states; the lowest norm among these states
occurs for those that transform in $(E'+\half, j_2'-\half, r'-1, R_j',R'_{m-2}+1)$
($j=2 \ldots m-2$); this norm is given by \eqn\normsL{\|~~\|^2_{R,sub}=E'-3 j_2
+3 \delta_{j_2 0} -3 -3 {\sum_{k=1}^{m-2} k R'_k \over m-1}
+{r'(4-m) \over (m-1)} \equiv E'+3 \delta_{j_2,0}-E_{2 sub}'(
j_2', r', R_i').}

Acting on the lowest weight states with supersymmetries
uncharged under $j_2$, we find a set of states; the lowest norm occur
for those states that transform in $(E'+1, j_2, r'+1, R'_{1}+1, R'_j)$
($j=2 \ldots m-2$). The norm of these states is given by
\eqn\normsR{\|~~\|^2_{sub,L}
=E'-{3 \over 2} {\sum_{k=1}^{m-2} (m-1-k) R'_k \over m-1}
-{r'(4-m) \over 2 (m-1)} \equiv E -E_{1 sub}'( r', R_i').}
Unitary representations occur when $E' \geq E'_{1 sub}$ and $E' \geq E'_{2 sub}$.
When these inequalities are strictly satisfied, the representations are
long and are denoted by $({\bf aa})_{sub}[E', j_2', r', R'_i]$. Representations
with $E'=E'_{2 sub}$ are short, and are denoted by $({\bf xc})_{sub}[j_2', r', R'_i]$.
When $j_2'=0$ the null states of
this representation occur at level 2. In addition, at $j_2'=0$ we
have a short representation at $E'=E'_{2sub}-3$, denoted by $({\bf xb})_{sub}[0, r',
R'_i]$. Representations with $E'=E_{1sub}'$ are denoted by $({\bf bx})_{sub}[j_2',r', R'_i]$.

Now consider a representation $R$ of the full algebra that is of the form
${\bf cx}$.
The highest weight state of $R$ has $\Delta = 2$. Acting on this with $Q' = Q^{+\half , 1}$,
we obtain a state with charges $(E+\half , J_1+\half ,J_2,r+1,R_1+1,R_i)$.
This state has $\Delta = 0$ and serves as the highest weight of the
representation $R^{'}$ of the subalgebra.
If $R$ is of the form ${\bf b x}$ then,
its highest weight state has $\Delta = 0$ and also serves as the highest weight
of $R^{'}$. If $R$ is of the form ${\bf a x}$, then it has no states with $\Delta = 0$.

Let us investigate if the representation $R'$ so obtained satisfied the
unitarity bounds from \normsL.

First, consider the case where $R$ is ${\bf c x}$.
Then highest weight of $R'$ is specified by the charges given by $\vec c$ in \defqbc .
Substituting these values of the charges into
equation \normsL, \normsR\ we find that
\eqn\normsimple{\eqalign{
\|~~\|^2_2 &= 3 \delta^{j_2}_{0} + {3 \over 2}  (E_1 - E_2) \cr
\|~~\|^2_1  &= 3j_1 + 3 + { 3 \over 2}  R_1}}
So, $R'$ is long unless $(E_1 = E_2)$.
In this case, $R \sim {\bf c c}$ and $R'$ is short. If $j_2 = 0$ it is possible to have
$E_2 = E_1 + 2$, and then $R \sim {\bf c b}$ and $R'$ is short.

Now, let $R$ = ${\bf b x}$. Then \eqn\normbsimple{\eqalign{
\|~~\|_2^2 &= 3\delta^{j2}_{0} - 3 + {3 \over 2}(E_1 - E_2) \cr
\|~~\|_1^2 &= {3 \over 2} R_1
}}

If ${\bf x}$ is  ${\bf a}$ or ${\bf c}$, we  have $E_1 - E_2 \geq 2$.
 If this inequality is saturated, $R \sim {\bf b c}$ and $R'$ is short.
 $R'$ may also be short if $R_1 = 0$. Finally, when $j_2 = 0$
 and $E_1 = E_2$, $R \sim {\bf b b}$ and $R'$ is short.

Using all of this, the decomposition of long representations as they hit
unitarity bound follows immediately; we will not explicitly list the
character formulae.

\appendix{C} {Conventions and Computations for the $\CN=4$ Index}

\subsec{Weights of the Supercharges}
In this subsection we list the weights of the supercharges under the Cartan elements
$(E, J_1^3, J_2^3,R_1,R_2,R_3)$.

\eqn\Qcharges{ \eqalign{
Q^{1}_{\pm} &\rightarrow \{{1 \over 2},\pm {1 \over 2},0,1,0,0\}
\cr Q^{2}_{\pm} &\rightarrow \{{1 \over 2},\pm {1 \over
2},0,-1,1,0\} \cr Q^{3}_{\pm} &\rightarrow \{{1 \over 2},\pm {1
\over 2},0,0,-1,1\} \cr Q^{4}_{\pm} &\rightarrow \{{1 \over 2},\pm
{1 \over 2},0,0,0,-1\} \cr \bar{Q}_{1,\pm } &\rightarrow \{{1 \over
2},0,\pm {1 \over 2},-1,0,0\} \cr \bar{Q}_{2,\pm} &\rightarrow \{{1
\over 2},0,\pm {1 \over 2},1,-1,0\} \cr \bar{Q}_{3,\pm} &\rightarrow
\{{1 \over 2},0,\pm {1 \over 2},0,1,-1\} \cr \bar{Q}_{4,\pm}
&\rightarrow \{{1 \over 2},0,\pm {1 \over 2},0,0,1\} \cr } }

\subsec{Racah Speiser Algorithm}
\subseclab\racahspeiser
The Racah Speiser algorithm is an efficient way to calculate tensor products.
Consider a highest weight state $|\Lambda>$ and the complete set of states
in another representation $|\lambda_i>$. We denote the half sum of positive roots by $\rho$.

The Racah-Speiser algorithm tells us that to obtain the representations in the
tensor product, we need to perform the following two steps.
\item{1.} First count all representations $|\Lambda + \lambda_i>$,
 where $\Lambda + \lambda_i$ is in the fundamental Weyl
 Chamber[All weights are non-negative].\item{2.}
 If $\Lambda + \lambda_i + \rho$ is on the boundary of the Weyl chamber
 i.e. at least one weight is zero, then throw away this representation.
 ($\rho$ is the half-sum of the positive roots).
\item{3.}If $\Lambda + \lambda_i$ is not on the boundary of  the Weyl Chamber,
there exists a unique Weyl reflection, $\sigma$ such that
$\sigma(\Lambda+\lambda_i+\rho)-\rho$ is in the Weyl Chamber.
 Count this representation with a plus or a minus sign depending on the sign of $\sigma$.

We use this algorithm to obtain the state contents tabulated below. It is interesting that for the Yang Mills multiplet,
using the Racah Speiser algorithm automatically gives us the representations corresponding to the equations of motion with negative signs.

\subsec{State content of 'graviton' representations}
As explained in section \strcoup\  the spectrum of Type $II B$ supergravity compactified on $AdS_5\times S^5$ organizes into representations of the superconformal
algebra that are are built on a lowest weight
state that is a scalar in the $(n,0,0)_{SO(6)} = (0,n,0)_{SU(4)}$
representation of the R-symmetry group.
When restricted to $\Delta = 0$, they yield a short representation of the subalgebra that we shall call $S_n$. $S_n$ has
lowest weight $[E=n, j_2 = 0, R_2 =n, R_3 = 0]$. We may explicitly compute the $SU(2,1) \times SU(3)$ content of $S_n$ by starting with the lowest weight state, repeatedly acting on it with the $Q$ and $\bar{Q}$ operators, and deleting states of zero norm. This process is expedited by using the Racah Speiser algorithm explained in \racahspeiser.

In the table below we explicitly list the $SU(2,1) \times SU(3)$ content of $S_n$
using the notation $[E', j_2', R'_1, R'_2]$ where $[E',j'_2]$ specify the weight of
the lowest weight state under the compact $U(1) \times SU(2)$ subgroup of $SU(2,1)$
and $[R'_2,R'_3]$ are Dynkin labels for $SU(3)$.
This can also be found by looking at the list of Kaluza Klein modes in \gunmar .

\bigskip{
\leftline{\bf Table 3: Content of $S_n$}
\vglue .5\baselineskip
\offinterlineskip
\def\tablerule{\noalign{\hrule}}
\halign{
\tabskip = .7em plus 1em
\hfil\strut #  \hfil \vrule &\hfil\strut # \hfil \vrule &\hfil\strut # \hfil \vrule &\hfil\strut # \hfil  \cr
\hfil ${\bf (-1)^F \,  E'}$  \hfil & \hfil ${\bf J'_2}$ \hfil & \hfil ${\bf R'_1}$ \hfil & \hfil ${\bf R'_2}$ \hfil\cr
\noalign{\hrule}
 $n$  &  0& n  & 0  \cr
 $-(n+{1 \over 2})$  & ${1\over 2}$ &  n-1  &  0    \cr
 $n+1$ & 0 &  n-2  &  0   \cr
 $-(n+1)$  &0 & n-1  &  1   \cr
 $n+{3\over 2}$ & ${1 \over 2}$  &  n-2  & 1  \cr
 $-(n+2)$  & 0 &  n-3  &  1  \cr
 $n+2$  & 0 &  n-1  &  0   \cr
 $-(n+{5 \over 2})$ & ${1 \over 2}$ & n-2  &  0  \cr
 $n+3$  & 0 & n-3  &  0  \cr
}}
\bigskip
For $n=2$ we just drop the lines containing $n-3$.

On the other hand, for $n=1$ we have further shortening and we find

\bigskip{
\leftline{\bf Table 4: Content of $S_1$}
\vglue .5\baselineskip
\offinterlineskip
\def\tablerule{\noalign{\hrule}}
\halign{
\tabskip = .7em plus 1em
\hfil\strut #  \hfil \vrule &\hfil\strut # \hfil \vrule &\hfil\strut # \hfil \vrule &\hfil\strut # \hfil  \cr
\hfil ${\bf (-1)^F \, E'}$  \hfil & \hfil ${\bf J_2'}$ \hfil & \hfil ${\bf R_1'}$ \hfil & \hfil ${\bf R_2'}$ \hfil\cr
\noalign{\hrule}
1 & 0 & 1 & 0  \cr
-$3 \over 2$ & ${1 \over 2}$ &0 & 0 \cr
-2 &0 &0 &1 \cr
3 & 0 &0 &0 \cr
3 & 0 &0 &0 \foot{This term comes from the fermionic equation of motion, hence it counts with a positive sign} \cr
}
}

\subsec{Character of $SU(3)$}
We wish to compute the quantity
\eqn\suthreechar{
\chi_{\rm R}(\theta_1, \theta_2) = {\rm Tr}_{\rm R}\exp{i (R_1 \theta_1 + R_2 \theta_2)},
}
where $R_1$ and $R_2$ form the Cartan subalgebra of $SU(3)$. We denote the eigenvalues of the highest weight state of a representation, under the operators $R_i$, by $r_i$. Furthermore, we define $v_1 = \exp{- i \theta_2}$, $v_2 = \exp{i \theta_1}$, $v_3 = \exp{i (\theta_2 - \theta_1)}$ \foot{These are the weights of the fundamental representation of $SU(3)$}

Then, the character \suthreechar\ is given by the Weyl Character Formula \FuchsJV.
\eqn\suthreecharacter
{
\chi_{R_1,R_2} = {\left|\matrix{v_1^{R_1+1} & v_2^{R_1+1} & v_3^{R_1+1}\cr v_1^{-R_2-1} & v_2^{-R_2-1} & v_3^{-R_2-1}\cr 1&1&1}\right| \over \left|\matrix{v_1^{1} & v_2^{1} & v_3^{1}\cr v_1^{-1} & v_2^{-1} & v_3^{-1}\cr 1&1&1}\right| }}

\subsec{Translation between bases}

$SO(6) \rightarrow SU(4)$

First, we show how to translate between $SO(6)$ and $SU(4)$ notation. Denote the $SO(6)$ Dynkin labels by $q_1, q_2, q_3$ and the $SU(4)$ Dynkin labels by $R_1,R_2,R_3$.

\eqn\transfoursix{
\eqalign{
q_1 &= {R_1 \over 2} + R_2 + {R_3 \over 2} \cr
q_2 &= {R_1 \over 2} + {R_3 \over 2} \cr
q_3 &= {R_1 \over 2} - {R_3 \over 2}
}}

$H',R'_i \rightarrow L_i$

Next, we show how to translate between the basis formed by the $L_{i}, J_2$ and the Cartan generators of the subalgebra $H+J_1,J_2,R'_1,R'_2$ defined above. Note that $R'_1 = R_2$, $R'_2 = R_3$.
Moreover, recall that the $L_i$ are specified by \newvar\ which we recapitulate here:
\eqn\newvarrecap{L_1 = E + q_1 - q_2 - q_3,~L_2 = E + q_2 - q_1 - q_3,~L_3 = E + q_3 - q_1 - q_2}
Under the condition $\Delta = 0$ we find (denoting $H+J_1 = H'$)
\eqn\transLA{
\eqalign{
L_1 &= {2 \over 3} H' + {2 \over 3} ( 2R'_1 + R'_2) \cr
L_2 &= {2 \over 3} H' + {2 \over 3} ( -R'_1 + R'_2) \cr
L_3 &= {2 \over 3} H' + {2 \over 3} ( -R'_1 - 2R' _2)}}

In turn this implies a relationship between the chemical potentials.
If $\theta_{H'} H' + \theta_1 R'_1 + \theta_2 R'_2 =
 \gamma_1 L_1 + \gamma_2 L_2 + \gamma_3 L_3$ then,
\eqn\transchem{
\eqalign{
\theta_{H'} &= {2 \over 3}(\gamma_1 + \gamma_2 + \gamma_3) \cr
\theta_1 &= {2 \over 3} (2 \gamma_1 - \gamma_2 - \gamma_3) \cr
\theta_2 &= {2 \over 3}( \gamma_1 + \gamma_2  - 2 \gamma_3) }}

\subsec{Index on the Fock space} {
Let us say that we have the single particle index
\eqn\singlepart{
Z_{sp} = \sum_i x^B_i - \sum_i x^F_i
}
where the index $i$ runs over all the bosons and all the fermions.
Then the index for a multiparticle system is given by
\eqn\multip{
Z_{Fock} = \prod_i { (1 - x_i^F) \over (1- x_i^B) } =
e^{ \sum_{n}{ 1 \over n}  Z_{sp}( x^n) }
}
So we find \sugraindex .

\appendix{D}{Comparison of the Cohomological Partition Function and the
Index}

Let the number of states with charges $J_1, J_2, L_i$ be given  by
$e^{S(J_1, J_2, L_i)}$. Then \eqn\indsum{ \eqalign{ Z_{\rm
free}&=\sum_{J_1, J_2, L_i} \exp\left[{S(J_1, J_2, L_i)-\sum_i
\gamma_i L_i -{ 2 \zeta J_2} } \right] \cr {\cal I}_{YM}^{WL}
&=\sum_{J_1, J_2, L_i} \exp\left[{S(J_1, J_2, L_i)-\sum_i \gamma_i
L_i -{2 \zeta J_2} } \right](-1)^{2 (J_1+J_2)} }}where we have
set all chemical potentials that couple to charges outside
$SU(2,1|3)$ to zero in $Z_{\rm free}$. Let \eqn\intermsum{\exp
\left[ N^2 S_{\rm eff}(\tilde j_1, \gamma_i) \right]=\sum_{ J_2,
L_i} \exp\left[S(J_1, J_2, L_i)-\sum_i \gamma_i L_i -{2 \zeta
J_2} \right]  . }
 where  $\tilde j_1 \equiv J_1/N^2 \gg 1$ and
$\gamma_i \ll 1$. Let us assume that
 $S_{\rm eff}$ is independent of $N$ in the large
$N$ limit. We certainly have this property in the free theory, and
we expect it in the interacting ${\cal N}=4$ theory,  but it does
not have to hold for every theory. We can then rewrite \indsum\  as
\eqn\indsumo{ \eqalign{ Z_{\rm free} &=\sum_{J_1}
\exp\left[N^2S_{\rm eff}(\tilde j_1, \zeta, \gamma_i) \right] \cr
{\cal I}_{YM}^{WL} &=\sum_{J_1} \exp \left[N^2 \left\{S_{\rm eff}(\tilde
j_1, \zeta+\pi i, \gamma_i) + 2 i \pi \tilde j_1 \right\} \right]}}
 Let us assume that that at fixed values of
$\zeta, \gamma_i$ has a maximum at $\tilde{j} = a(\theta,
\gamma_i) $ and that \eqn\saddle{\eqalign{ S_{\rm eff} (a + \delta,
\zeta, \gamma_i) &\approx S_0 - 2 b^2 \delta^2 \cr S_0 &= S_{\rm
eff}(a, \zeta, \gamma_i). \cr} } The contribution of this saddle
point to the  partition function in the first line of \indsumo\ is
easily estimated\foot{For instance one could convert the sum into an
integral using the Euler McLaurin formula \arfken\ and approximate
the integral using saddle points. A more careful estimate may be
obtained by Poisson resumming, see the next paragraph.} by
\eqn\pf{Z_{\rm free} \approx \sqrt{2 \pi \over b^2 N^2}\exp\left[N^2
S_0\right].}

An estimation of the Index in the second line of \indsumo\ is a more
delicate task as the summand changes by large values over integer
spacings. To proceed we will {\it assume} that $S_{\rm eff}(j_1,
\zeta, \gamma_i)$ is a continuous function; i.e. that it does not
evaluate to discontinuously different answers for integral and half
integral values of $J_1$. This is a nontrivial assumption, which we
believe to be true for free Yang Mills theory, but will not always
be true in every theory.  Under this assumption we will now estimate
the contribution of the saddle point at ${\tilde j_1= a}$ to the
index by \eqn\indexest{\eqalign{{\cal I}_{YM}^{WL} &=e^{N^2
S_0}\sum_{m=-\infty}^\infty \exp \left[-{b^2 m^2 \over 2 N^2} +\pi i
m \right] \cr &= e^{N^2 S_0} \sum_{k=-\infty}^\infty \sqrt{2 \pi
\over b^2 N^2} \exp \left[ {N^2 (2 \pi)^2 \over 2 b^2} (k-\half)^2
\right] \cr &\approx 2 \sqrt{2 \pi \over b^2 N^2} \exp\left[ N^2(S_0
-{\pi^2 \over 2 b^2}) \right]  }} where we have used the Poisson
resummation formula to go from the first to the second line of
\indexest.

Note that the contribution of the saddle point at ${\tilde j_1}=a$
to the index is supressed compared to its contribution to the
partition function. Moreover, if $S_0<\pi^2 /2b^2$, the contribution
of this saddle point is formally of order $e^{-a N^2}$; which means
that the neighborhood of the saddle point does not contribute
significantly to the index in the large $N$ limit; the Index
receives its dominant contributions from other regions of the
summation domain. A estimation from formulas of \pfrel, \en\ puts us
in this regime

As a toy example of the suppression described in the last two
paragraphs, consider the two identities
\eqn\toy{\eqalign{Z=(2+1)^N&=\sum_k  2^k {N! \over k! (N-k)!} \cr
I=(2-1)^N&=\sum_k  2^k {N! \over k! (N-k)!} (-1)^{N-k}.}} The
summation over $k$ in the first of \toy\ may be approximated by the
integral \eqn\intapr{ \int_{x=0}^1 e^{N \ln {2^x \over x^x
(1-x)^{1-x} }}, } which localizes around the saddle point value
$x^s={2\over 3}$ at large $N$, yielding  $Z=3^N$. The contribution
to $I$ from this saddle point, on the other hand, is proportional to
$e^{N( \ln 3 -{\pi^2 \over 3})}$, and so is utterly negligible,
consistent with the fact that $I$  evaluates to unity.
\foot{Actually, a computation very similar to this toy example
explains why the index grows more slowly that exponentially with
energy in the `low temperature phase' (while the cohomological
partition function displays exponential growth in the same phase).
The number of states that contribute at energy $E$ to the index is
given by the coefficient of $x^E$ in \ymindex. This is given by a
multinomial expansion. When we weight the sum with $(-1)^F$, the
multinomial sum stops growing exponentially just like \toy\ above.
Hence, the Index never goes through a Hagedorn like transition.}

\vfil

\listrefs

\end{document}